\documentclass[sigconf, acmpreprint]{acmart}
\usepackage[font=small, labelformat=simple]{subcaption}
\usepackage[font=small, labelformat=simple]{caption}
\usepackage{cleveref}
\usepackage{enumitem}
\usepackage{color}
\usepackage{multicol}
\usepackage{multirow}
\usepackage{array}
\usepackage{color, colortbl}
\usepackage{booktabs}
\usepackage{rotating}
\usepackage{stmaryrd}
\usepackage{amsthm}
\usepackage{soul}
\usepackage{xcolor}
\usepackage{soul}
\usepackage{scalerel,graphicx,xparse}
\usepackage{xspace}
\usepackage{microtype}
\usepackage{hyperref}
\usepackage{url}
\usepackage{booktabs}
\usepackage{graphicx}
\usepackage{fancybox}

\usepackage[ruled,lined,noend]{algorithm2e}
\usepackage{amsmath}

\newcommand{\tool}{\textsc{Attune}\xspace}

\newcommand{\numformative}[0]{12}

\newcommand{\ttt}[1]{{\small \texttt{#1}}\xspace}
\newcommand{\topic}[1]{\smallskip \noindent{\bf #1.}}

\newcommand{\bhavya}[1]{}
\newcommand{\meng}[1]{}
\newcommand{\shreya}[1]{}
\newcommand{\sep}[1]{}
\newcommand{\aditya}[1]{}
\newcommand{\bjoern}[1]{}
\newcommand{\rebecca}[1]{}

\definecolor{teal}{RGB}{17, 94, 89}
\definecolor{lightteal}{RGB}{240, 253, 250}

\newcommand{\tealbox}[1]{%
  \par\smallskip\noindent
  \begingroup
    \setlength{\fboxrule}{0.1pt}
    \setlength{\fboxsep}{1mm}%
    \fcolorbox{teal}{lightteal}{%
      \parbox{\dimexpr\linewidth-2\fboxrule-2\fboxsep\relax}{\small #1}}%
  \endgroup
  \par\smallskip                   
}

\newcommand{\del}[1]{}
\newcommand{\add}[1]{#1}

\AtBeginDocument{%
  }

\copyrightyear{2026}
\acmYear{2026}
\setcopyright{cc}
\setcctype{by}
\acmConference[UIST '26]{The 39th Annual ACM Symposium on User Interface Software and Technology}{November 02--05, 2026}{Detroit, MI, USA}
\acmBooktitle{The 39th Annual ACM Symposium on User Interface Software and Technology (UIST '26), November 02--05, 2026, Detroit, MI, USA}
\acmDOI{10.1145/3830398.3830567}
\acmISBN{979-8-4007-2856-3/2026/11}

\begin{document}

\title{Who's Keeping Score? Interactive Steering of LLM-Powered Scoring with \tool{}}

\author{Bhavya Chopra}
\email{bhavyachopra@berkeley.edu}
\affiliation{%
  \institution{UC Berkeley}
  \city{Berkeley}
  \state{CA}
  \country{USA}
}

\author{Meng Chen}
\email{meng.chen@berkeley.edu}
\affiliation{%
  \institution{UC Berkeley}
  \city{Berkeley}
  \state{CA}
  \country{USA}
}

\author{Rebecca Dang}
\email{rdang@berkeley.edu}
\affiliation{%
  \institution{UC Berkeley}
  \city{Berkeley}
  \state{CA}
  \country{USA}
}

\author{Chanbin Park}
\email{chanbin.park@berkeley.edu}
\affiliation{%
  \institution{UC Berkeley}
  \city{Berkeley}
  \state{CA}
  \country{USA}
}

\author{Shreya Shankar}
\email{shreyashankar@berkeley.edu}
\affiliation{%
  \institution{UC Berkeley}
  \city{Berkeley}
  \state{CA}
  \country{USA}
}

\author{Sepanta Zeighami}
\email{zeighami@berkeley.edu}
\affiliation{%
  \institution{UC Berkeley}
  \city{Berkeley}
  \state{CA}
  \country{USA}
}

\author{Bjoern Hartmann}
\email{bjoern@eecs.berkeley.edu}
\affiliation{%
  \institution{UC Berkeley}
  \city{Berkeley}
  \state{CA}
  \country{USA}
}

\author{Aditya Parameswaran}
\email{adityagp@berkeley.edu}
\affiliation{%
  \institution{UC Berkeley}
  \city{Berkeley}
  \state{CA}
  \country{USA}
}

\renewcommand{\shortauthors}{Chopra et al.}

\begin{teaserfigure}
    \centering
    \includegraphics[width=1\linewidth]{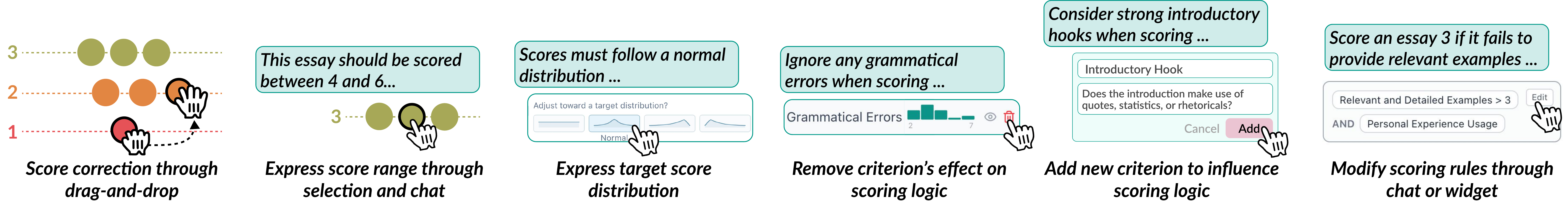}
    \caption{\tool offers {\em steering interactions} for LLM-powered scoring. Users can refine scoring logic through direct manipulation or natural language chat: correcting individual scores via drag-and-drop, expressing score ranges by referring to specific records, reshaping target distributions, and modifying scoring criteria and rules.}
    \Description{}
    \label{fig:teaser}
\end{teaserfigure}

\begin{abstract}

Large language models (LLMs) are increasingly used to {\em score} text records at scale (e.g., rating candidate resumes on a 1–5 scale). However, existing LLM-powered approaches do not account for the fact that effective scoring requires both holistic understanding of records and locally consistent judgments across similar ones. We present \tool, a mixed-initiative system for steerable LLM-powered scoring. Given a task description and scoring range, \tool performs pairwise comparisons across records to develop a global understanding first, and then resolves these comparisons into consistent score assignments—deriving scoring criteria and rules bottom-up in the process. These serve as shared representations of scoring logic that users can inspect and edit. Based on insights from a formative study ($n=12$), \tool's interface introduces novel steering interactions that allow users to deterministically refine scoring logic. Users can provide examples, directly edit criteria, rules, or target distributions, and give natural language feedback---with all refinements compiling into constraints that guide re-scoring. We validate our approach through a technical evaluation across three workloads and a user study with domain experts ($n=8$) in healthcare, law, education, and AI evaluation.

\end{abstract}


\maketitle

\section{Introduction} 
\label{sec:intro}

\begin{figure*}
    \centering
    \includegraphics[width=0.96\linewidth]{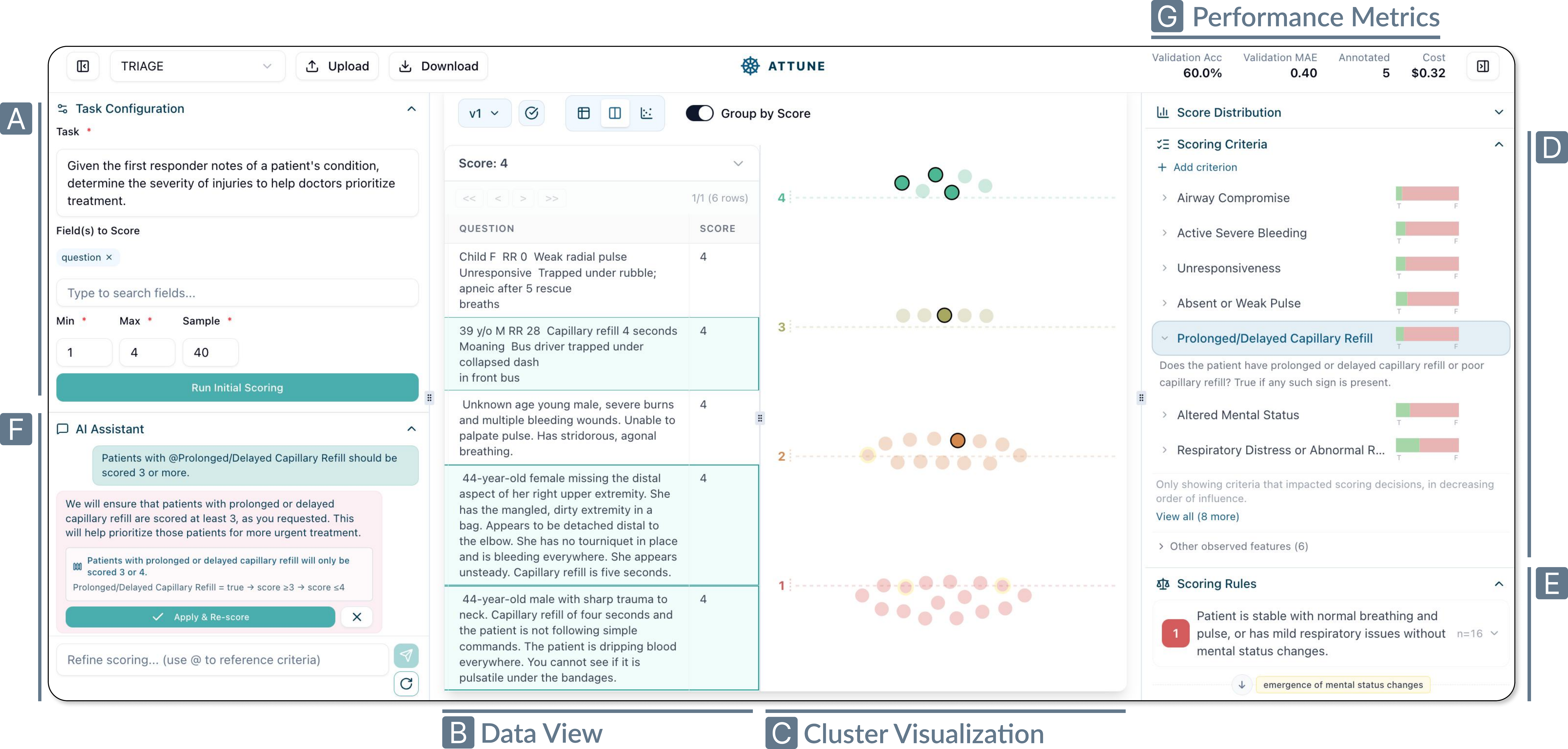}
    \caption{\tool{}'s Interface. Users begin by (A) specifying their scoring task, and inspecting assigned scores via the (B) Data View and (C) Cluster Visualization. Users inspect and manipulate scoring logic through (D) Scoring Criteria and (E) Scoring Rules. (F) An AI chat assistant translates all user provided feedback into steering actions to guide re-scoring. (G) Users track performance over a self-curated validation set.}
    \Description{}
    \label{fig:interface}
\end{figure*}

Large language models (LLMs) are increasingly used to score text records at scale~\cite{gu2025surveyllmasajudge}---assigning numerical ratings that drive downstream decisions in domains including clinical triage~\cite{gourabathina2025medium, croxford2025evaluating, hager2024evaluation}, grading~\cite{kim2024gpt}, screening job candidates~\cite{lo2025ai}, and peer reviewing~\cite{latona2024ai}. The canonical way practitioners achieve this is by writing a {\em scoring prompt} detailing the criteria, rubrics, and a scoring range (e.g., 1--5) to evaluate records individually or in batches.

\topic{Motivating Scenario}
Consider a clinician preparing for a mass casualty incident using an LLM to develop and validate triage scoring logic on simulated patient records, scoring injury severity on a 1–4 scale (4 = most severe). \add{The following example is taken from the \textsc{Triage} dataset~\cite{kirch2024triage}, where GPT 4.1 produces LLM scores, and clinicians provide ground-truth labels}:

\tealbox{
\textbf{Patient A:} \emph{6 y/o F. No breathing. No pulse. Unresponsive. Trapped under bus seat.}
\hfill LLM Score:\,\textbf{4/4} ~|~ Clinician:\,\textbf{4/4}\\[1pt]
\textbf{Patient B:} \emph{Child M. RR 36. Pulse present. Makes eye contact. Bleeding from ears, bruise on neck.}
\hfill LLM Score:\,\textbf{4/4} ~|~ Clinician:\,\textbf{2/4}
}

\noindent The LLM assigns identical scores to these patients with vastly different prognoses. The clinician cannot determine {\em why} the model equated these cases---is it ignoring respiratory rate? Overweighting {\em ``bleeding from ears?''} Even if they could diagnose the problem, they have no mechanism to fix it deterministically. Without the ability to assess or reliably manipulate the scoring logic, the clinician must either accept opaque outputs or abandon automation entirely. 
\qed{}

\smallskip

\noindent We identify three challenges with prompt-based scoring approaches that make them unreliable and difficult to control:

\topic{First, lack of holistic data understanding} Scoring is inherently comparative. For example, whether a patient can be prioritized soon depends on other patients awaiting treatment. Prompt-based scoring treats records in isolation, producing {\em inconsistent} judgments, where meaningfully different records may receive identical scores, while comparable records get scored differently~\cite{ouyang2022training, rafailov2023direct, sahoo2025quantitative, liu2025pairs, zhao2025llmorderby}.  

\topic{Second, lack of interpretability} It is not clear {\em why} an LLM scores certain records higher or lower than others. One could generate post-hoc explanations (such as chain-of-thought traces) with LLMs, but this must be done on a per-record basis, making it challenging to aggregate explanations at scale and make sense of overall scoring logic. Further, such explanations may even lead users to over-trust model outputs when they do not faithfully reflect implicit behavior~\cite{bansal2021does, vaccaro2024combinations}. Existing tools for LLM evaluation help users audit outputs---through side-by-side comparisons~\cite{Kim_2024} or fragmentation against user-defined criteria~\cite{kim2025evalet}---but the underlying scoring logic remains opaque. 

\topic{Third, lack of steerability} The relationship between a natural language prompt and the scores it produces is indirect and brittle: small edits can shift score distributions unpredictably, while substantive changes can sometimes have no noticable effect~\cite{murugadoss2025evaluating}. Users struggle to convey and enforce desired behavior through natural language instructions~\cite{zamfirescu2023johnny}, and face what~\citet{subramonyam2024bridging} call the {\em gulf of envisioning} between their goals and effective prompt formulations. An added challenge is that users often themselves do not know what the scoring criteria should be upfront---they discover and iterate on criteria through engaging with LLM outputs---a phenomenon~\citet{shankar2024validates} term {\em criteria drift}. 
Together, these challenges produce scoring decisions whose underlying rationale users can neither inspect nor manipulate---leaving them unable to specify, comprehend, or act on scoring logic. 

To address these challenges, we present \textbf{\tool{}, a mixed-initiative LLM-powered scoring system}. \tool introduces a novel scoring algorithm (\Cref{fig:scoring_algorithm}) that performs pairwise comparisons across records to develop holistic data understanding rather than assessing them in isolation. It then resolves these  judgments to assign {\em consistent \add{and relative}} scores, and derives scoring criteria {\em bottom-up} in the process (as opposed to relying on users to specify criteria or rules upfront). \add{\tool then learns scoring rules retrospectively from resulting score assignments and criteria, and presents them as interpretable summaries of scoring logic.}

Informed by a probe-driven formative study ($n=12$) that revealed practitioners' refinement strategies, \tool{} introduces novel \textbf{{\em steering interactions}} (\Cref{fig:teaser}): users can directly manipulate scoring criteria (\Cref{fig:interface}D) and rules (\Cref{fig:interface}E) as shared representations of scoring logic, provide examples, modify target distributions, or delegate refinements to a chat assistant that automates steering with human oversight (\Cref{fig:interface}F). All steering interactions compile into structured constraints that deterministically guide re-scoring (\Cref{tab:steering}). \add{\tool is intended to add human oversight to LLM-based scoring by surfacing criteria and rules as interpretable patterns and summaries. With increasing use of opaque LLM-based scoring across sensitive and safety-critical domains, \tool attempts to offer a calibrated second opinion to catch inadvertent errors by experts. We do not advocate for LLM-based scoring to entirely replace human judgment and oversight, and we reflect on the scope and appropriate use of \tool and its steering interactions in \Cref{sec:discussion}.}

Our technical evaluation across three workloads shows that \tool improves accuracy and steerability over prompt-based scoring (§\ref{sec:tech-eval}). A user study with domain experts ($n = 8$) in healthcare, law, education, and AI evaluation revealed that participants trusted bottom-up criteria quickly, composed steering interactions fluidly, and grounded confidence in the interpretability of scoring logic (§\ref{sec:findings}).
In summary, we contribute:
\begin{itemize}[nosep, leftmargin=*]
    \item Identification of challenges with prompt-based scoring---lack of holistic understanding, interpretability, and steerability---motivating the need for shared representations of scoring logic.
    \item Findings from a probe-driven formative study ($n=12$) informing the space of steering interactions for LLM-powered scoring.
    \item The design and implementation of \tool, a mixed-initiative system for LLM-powered scoring with a novel comparative scoring algorithm and interface for interactive steering.
    \item Findings from a technical evaluation across three workloads and a user study with domain experts ($n=8$) validating our approach.
\end{itemize}
\section{Related Work}

We review challenges with LLM-powered scoring, current top-down and bottom-up approaches toward LLM-powered evaluation, and principles from mixed-initiative interfaces that inform \tool{}'s design and implementation.

\subsection{Challenges with LLM-powered Scoring}

\topic{Known issues with LLM evaluators} 
A growing body of work has documented systematic shortcomings of LLM-based evaluation.
LLM judges exhibit sensitivity to surface-level features such as input ordering and verbosity, and have self-preference bias~\cite{zheng2023_mtbench_chatbotarena, wataoka2024self, saito2023verbosity}. Rubric-based approaches suffer from poor adherence, as LLMs frequently deviate from specifications in unpredictable ways~\cite{murugadoss2025evaluating, wang2025trustjudge}, and make biased judgments despite meticulously specified criteria~\cite{li2024llms}. In high-stakes domains, these issues carry real consequences:~\citet{hager2024evaluation} show that LLMs prioritize patients differently based on formal versus informal symptom descriptions, while~\citet{hu2026evaluation} demonstrate that flipping demographic attributes can shift sentencing recommendations. Studies in healthcare~\cite{zewe2025llms_unrelated_medical_treatments, gourabathina2025medium, suzgun2024beliefmachineinvestigatingepistemological} and law~\cite{rozado2026fairness, hu2026evaluation} conclude that LLM judgments require substantial expert auditing before deployment.

\topic{Consistency in scoring} A persistent challenge in scoring is inconsistency: meaningfully different inputs may receive identical scores, while comparable inputs get scored differently. Multiple papers have shown that relative judgments (such as pairwise comparisons or preference-based evaluations) yield more consistent orderings than absolute scoring~\cite{ouyang2022training, rafailov2023direct, 
sahoo2025quantitative, liu2025pairs}. Recent data processing systems even offer a rank operator~\cite{shankar2025docetl, patel2025lotus, zhao2025llmorderby}, demonstrating the value of comparative evaluation.
However, these approaches offer no mechanism for human oversight. 
\tool{} applies the insight that relative judgments improve LLM-powered scoring consistency, while introducing scoring criteria and rules as shared representations for interactive steering.

\topic{Interaction challenges in LLM-mediated workflows}
More broadly, recent work uncovers interaction challenges when users attempt to steer LLM-mediated systems. \citet{zamfirescu2023johnny} show that end-users struggle to convey desired behavior through NL instructions and often abandon systematic evaluation. For scoring, this challenge is acute: the mapping from NL prompts to score assignments is indirect and opaque, leaving users unable to anticipate how prompt edits will propagate. \citet{subramonyam2024bridging} formalize this issue as the gulf of envisioning---the cognitive distance between a user's goal and an effective prompt formulation---which is exacerbated in scoring because stochasticity in LLM reasoning makes outcomes across several items unpredictable.
\citet{shankar2025docwrangler} also present three gulfs for LLM-powered data processing that are pertinent to scoring: the gulf of specification (difficulty encoding logic in prose), comprehension (limited user bandwidth for understanding all inputs), and generalization (uncertainty about prompts holding across the data). Van der Maden et al.~\cite{van2026results} identify a complementary results-actionability gap: practitioners struggle to translate LLM evaluation results into targeted interventions because scoring logic is not legible enough. For LLM-powered scoring, all of these gulfs operate simultaneously, leaving users unable to specify, comprehend, or act on the scoring logic.

\subsection{Approaches to Audit \& Improve LLM-powered Evaluation}
\topic{Top-down approaches and their limitations}
Several interfaces support auditing LLM evaluations through {\em top-down workflows} that assume users can articulate criteria upfront: EvalLM~\cite{Kim_2024} provides side-by-side LLM output comparisons on user-defined dimensions. ChainForge~\cite{arawjo2024chainforge} and ChainBuddy~\cite{zhang2024chainbuddy} offer visual dataflow programming for hypothesis testing. Evalet~\cite{kim2025evalet} fragments LLM outputs into functional units for granular inspection against user-defined criteria. However, all of these tools require users to specify criteria upfront. In practice, users discover and iterate on criteria through engaging with LLM outputs---a phenomenon \citet{shankar2024validates} term {\em criteria drift}. For scoring, top-down specification is particularly fragile due to the long tail of data characteristics.

\topic{Bottom-up approaches} 
Recent work demonstrates the value of bottom-up approaches that surface latent themes from unstructured text. LLooM~\cite{lam2024concept} synthesizes human-interpretable concepts from text. PolicyMaps~\cite{lam2025policy} composes bottom-up concepts into corrective policies for content moderation. EvalGen~\cite{shankar2024validates} applies a mixed-initiative approach to LLM-based evaluation, proposing candidate evaluation criteria from users' analysis of LLM outputs.

\tool{} draws on the insight that scoring criteria should be derived bottom-up, but differs in the demands of scoring. Bottom-up criteria discovery alone is insufficient: scoring requires comparative evaluation to identify which criteria are common within a score group and which differentiate across groups. These criteria must compose into rules that collectively account for every input---not just targeted subsets. This distinguishes scoring from adjacent problems like policy authoring (as in PolicyMaps), which focuses on identifying and correcting AI safety violations in generated text, 
and concept induction (as in LLooM), which surfaces thematic patterns without needing to resolve them into exhaustive, mutually discriminating score assignments. Moreover, as data grows more diverse, rule complexity increases: decision boundaries must account for interacting criteria, and rules must handle all edge cases. 

\subsection{Mixed-Initiative Interfaces \& Interactive ML}
A central challenge across LLM-powered evaluation pipelines is how users and automated agents should share control. Theory for direct manipulation argues that reducing cognitive distance between users' goals and system representations increases engagement~\cite{hutchins1986direct}. Principles of mixed-initiative interaction~\cite{horvitz1999principles} and the debate between direct manipulation and interface agents~\cite{shneiderman1997direct} refine this: effective human-AI collaboration requires legible, manipulable control surfaces even when the system acts autonomously.  
DirectGPT \cite{masson2024directgpt} demonstrates that these principles can successfully translate to LLM-powered interfaces by mapping pointing, selection, and drag-and-drop actions into engineered prompts for text, code, and image editing.
The interactive machine learning (IML) literature has offered foundational patterns for steering and debugging model behavior~\cite{amershi2014power, amershi2019guidelines}---systems like What-If Tool~\cite{wexler2019if} 
and Gestalt~\cite{patel2010gestalt} operationalize mixed-initiative and direct manipulation principles with models that offered well-defined feature spaces and deterministic inference. However, the shift to LLMs adds a dimension of difficulty that traditional IML systems did not confront: no guarantee of deterministic output, no feature importance to inspect, no stable decision boundary to manipulate.

\noindent \textbf{\del{Summary.}}
\del{Attune synthesizes these ideas to perform relative and consistent scoring, discover bottom-up criteria and rules to capture the long tail of data characteristics, offer legibility and steerability through direct manipulation, and provide AI-assisted automation of repetitive or tedious steering sequences with human oversight.}

\section{\tool: Design}
To understand {\em what} makes scoring legible, we built a functional chat-based design probe for users to iterate over LLM-generated scores and conducted a formative study ($n=\numformative$) to isolate design considerations for \tool{}.

\subsection{Design Probe and Formative Study}
\label{subsection:probe}

\begin{figure}[htp]
    \centering
    \includegraphics[width=\linewidth]{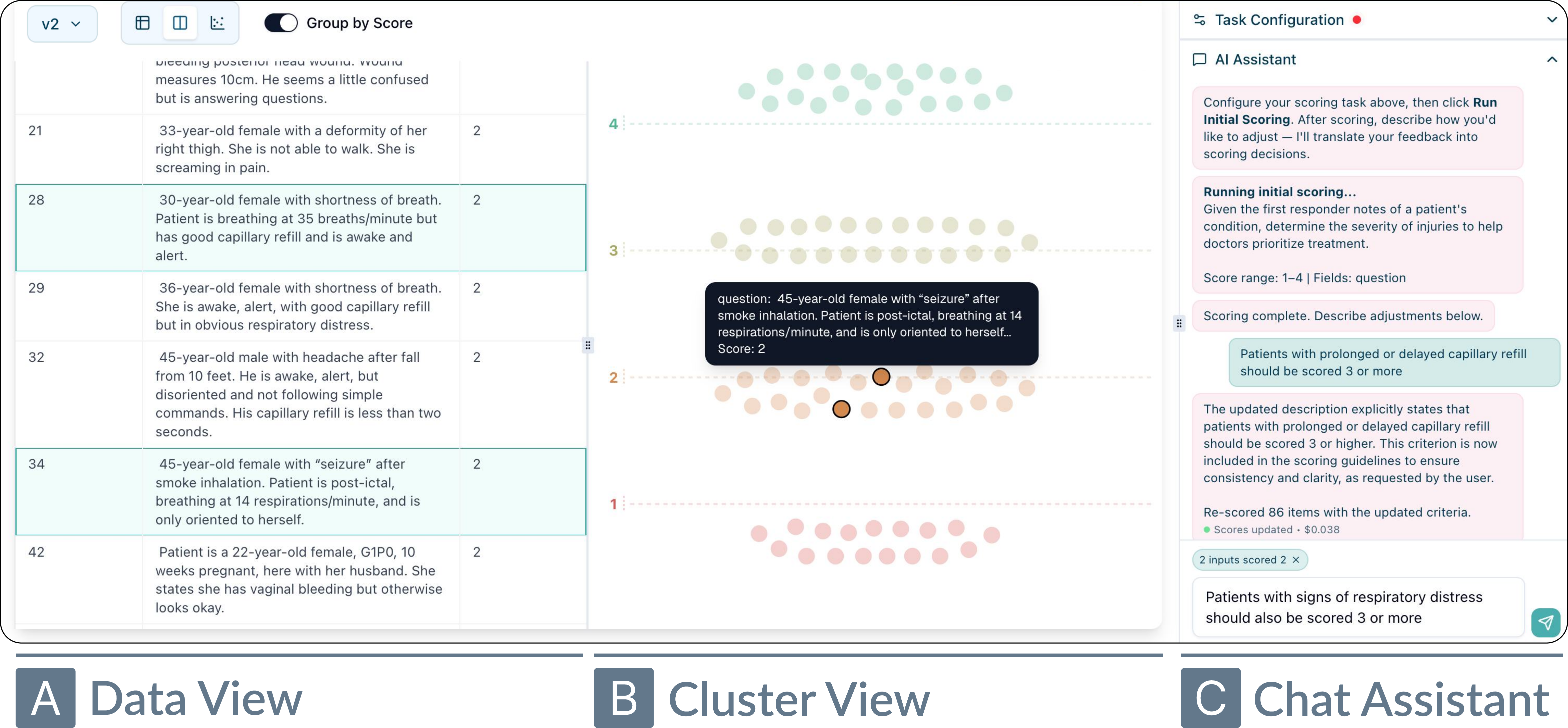}
    \caption{Screenshot of Study Probe}
    \Description{}
    \label{fig:probe}
\end{figure}

We implemented a functional probe as a web application\add{\footnote{\add{Available at} \textcolor{blue}{\url{https://attune-alpha.vercel.app/probe}}}} 
to observe how practitioners validate and refine LLM-generated scores via NL prompts and refinements.
The probe interface (Figure~\ref{fig:probe}) comprises a task configuration panel, a tabular data view (Figure~\ref{fig:probe}A), a cluster view grouping records by scores and linked bidirectionally with the table (Figure~\ref{fig:probe}B), and an AI assistant for NL refinement requests (Figure~\ref{fig:probe}C). Users could click on one or more rows in the data table to attach them as context for their chat message, enabling both record-specific feedback (e.g., ``fractures for ambulatory patients like \emph{these} should not be marked urgent'') and global feedback (e.g., ``prioritize patients with poor capillary refill''). For each refinement, the probe rewrites the scoring prompt based on the user's message and any selected context, then re-scores all records with the updated prompt. A version history dropdown and diff view support lightweight validation across iterations. Implementation details are provided in Appendix~\ref{appendix:probe}.

\noindent Using this probe, we conducted a formative study with $\numformative$ participants (F1--F12) recruited for their experience using LLM evaluators in enterprise pipelines. Each 45-minute remote session included a walkthrough of the probe followed by ~25 minutes of think-aloud interaction on one of four scoring tasks: patient triaging, essay scoring, monitor stand product relevance, and candidate resume screening---varying in expertise requirements (generic vs.\ domain-specific) and evaluation target (accuracy-based vs.\ subjective). Full study details are provided in Appendix \ref{appendix:formative}. 

\subsection{Findings and Design Considerations}
\label{subsection:considerations}

Our formative study surfaced challenges with prompting strategies and consistent feedback strategies used by experts. We identified four design considerations that inform \tool{}'s steering interactions and implementation.

\topic{C1: Ground scoring criteria and rules in data characteristics}
Participants consistently struggled to comprehend the full distribution of data characteristics, encountering the \emph{gulf of comprehension}~\cite{shankar2025docwrangler}. Because of this challenge, they discovered criteria incrementally through engagement with LLM outputs rather than arriving with a complete rubric upfront. F5 wondered, \emph{``are these criteria ever going to be exhaustive?''} and F10 noted the need for the system to \emph{``learn decision-making rules by breaking down the task into smaller verifiable criteria.''}
Other participants also echoed similar struggles (F1, F2, F4, F6, F7, F9, F11). These observations motivate a system that proactively derives criteria and rules bottom-up from data, as opposed to relying on users to provide evaluation criteria.

\begin{figure*}[htp]
    \centering
    \includegraphics[width=0.8\linewidth]{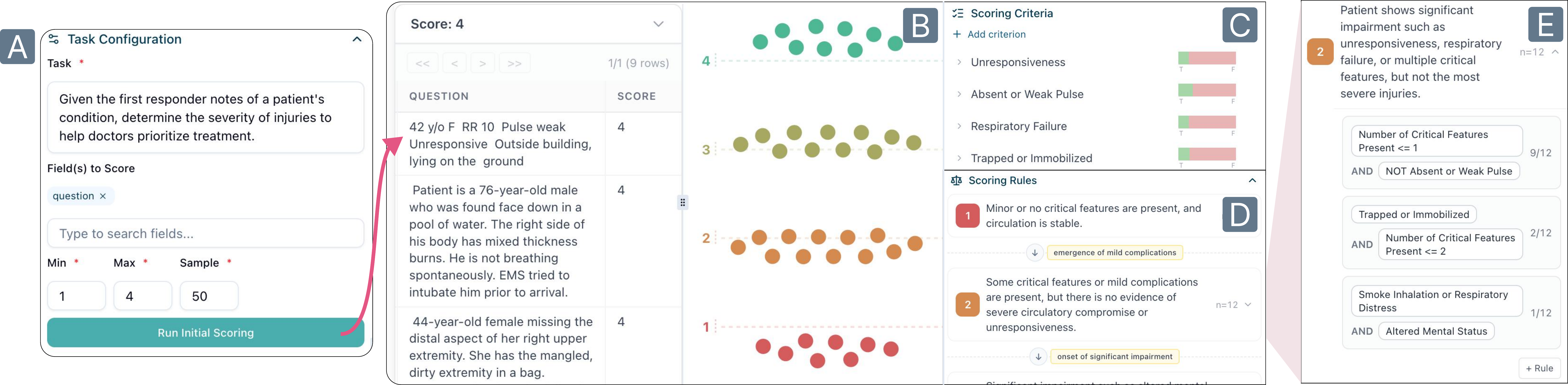}
    \caption{Initial Scoring. (A) User specifies their scoring task and clicks ``Run Initial Scoring.'' (B) User inspects assigned scores in the data table view and cluster visualization, (C) scoring criteria, and (D) scoring rules. (E) User clicks on a score group to inspect scoring rules in detail.}
    \Description{}
    \label{fig:init-run}
\end{figure*}

\topic{C2: Make scoring logic legible and manipulable at multiple granularity levels}
Participants needed explanations of scoring decisions at different levels---from individual items, to groups of items, to decision boundaries---and no single explanation sufficed. For example, while screening candidate resumes, F6 noticed that candidates scored 8 had ETL experience while those scored 9 or 10 did not, and wanted to understand which criteria the LLM prioritized. Participants also needed manipulability at specific granularities: F2 wanted \emph{``localized effects, [...] but the LLM re-scored everything,''} while F11 expected global effects from their refinements but found they \emph{``ended up touching very few candidates.''} These observations suggest that legibility and manipulability must operate across local (individual examples) and global (criteria, rules, decision boundaries, distributions) levels.

\topic{C3: Support direct manipulation and deterministic feedback mechanisms}
NL was frequently inadequate as a control mechanism. Participants struggled to formulate changes in prose (F4: \emph{``I don't know how to phrase this''}), and NL prompts produced unpredictable effects (F6: \emph{``Same prompt, same everything, but now I have a different distribution, why?!''}; F7: \emph{``the changes seem random''}). Several participants gravitated toward rule-like formulations---F5 wanted \emph{``deterministic tooling to say if the essay is less than 5000 characters don't score more than 5''}---but the probe offered no mechanism to enforce them. These observations argue for direct manipulation affordances and deterministic scoring rules.

\topic{C4: Support measuring progress toward scoring alignment}
Participants with access to ground truth had systematic validation strategies (F9 targeted mismatches; F10 started with the largest prediction gaps). Those with subjective tasks found it challenging to gauge progress---F1 did not know \emph{``how to track changes to scores for the resumes I have inspected in detail.''} These findings motivate support for building a personal ground-truth validation set and support tracking of progress.

\section{Feature Walkthrough of \tool}
\label{sec:walkthrough}

We illustrate \tool{}'s key features in continuation of the motivating scenario presented in \Cref{sec:intro}: a clinician using simulated patient records~\cite{kirch2024triage} from a mass casualty training exercise to develop and calibrate triage scoring logic. 
To begin, the clinician specifies their task, selects the patient status field (named \texttt{question}), 
and runs initial scoring (\Cref{fig:init-run}A).

\topic{Inspecting initial scores, criteria, and rules}
After initial scoring, the clinician reviews patients grouped by severity in the cluster visualization (\Cref{fig:init-run}B). To understand {\em why} patients landed where they did, they examine the scoring criteria (\Cref{fig:init-run}C) and rules (\Cref{fig:init-run}D) (\textbf{C1, C2}). Wanting a closer look at borderline cases, the clinician clicks on score group 3 (\Cref{fig:init-run}E), which reveals deterministic scoring rules (e.g., patients with fewer than 2 critical features, and not absent or weak pulse). 

\begin{figure}[htp]
    \centering
    \includegraphics[width=0.95\linewidth]{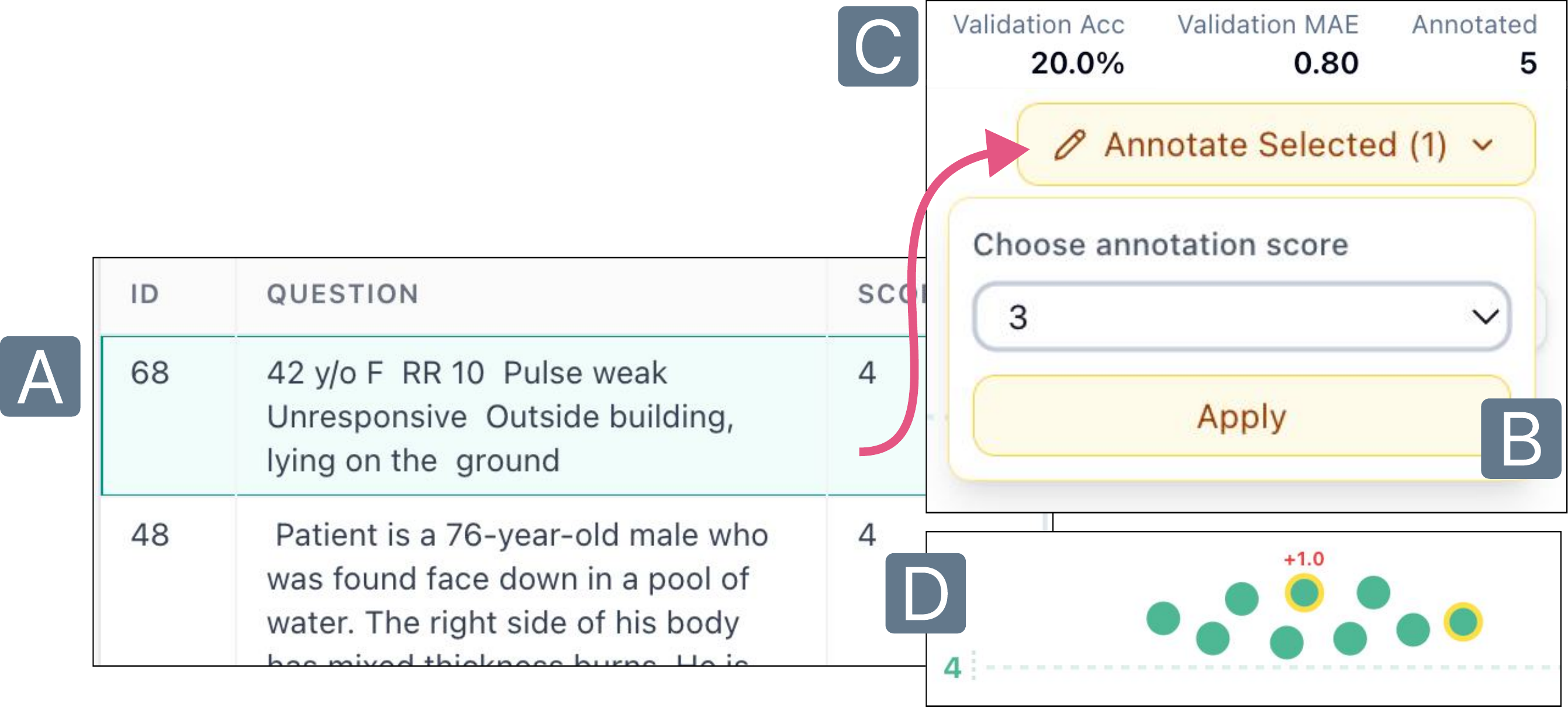}
    \caption{Annotating Samples. (A) User selects one or more inputs, and clicks (B) ``Annotate Selected'' to pick a ground-truth score. The user can (C) track accuracy and MAE for annotated samples, and (D) visually identify them with golden outlines to the inspect deviation between predicted and ground-truth scores.}
    \Description{}
    \label{fig:annotate}
\end{figure}

\topic{Annotating disagreements to seed a validation set}
While inspecting assigned scores, the clinician spots a patient scored 4 whom they believe should be a 3. Rather than immediately steering the system, they annotate the patient with their own ground-truth score (\Cref{fig:annotate}A--B), seeding a personal validation set. As annotations accumulate, \tool{} tracks accuracy and mean absolute error (MAE) against them (\Cref{fig:annotate}C), giving the clinician a concrete measure of alignment that persists across subsequent refinements (\textbf{C4}). The annotated samples now appear with golden outlines and visual indicators of deviation in the cluster visualization (\Cref{fig:annotate}D). 

\begin{figure}[htp]
    \centering
    \includegraphics[width=0.9\linewidth]{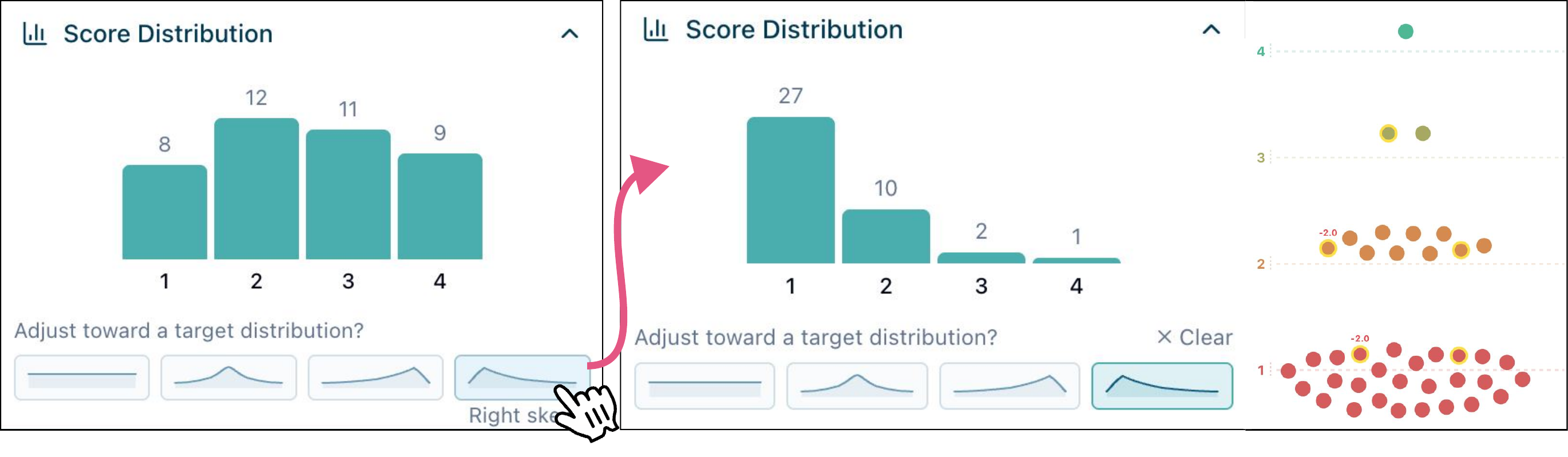}
    \caption{Reshaping Score Distribution. User selects ``Right skew'' as the target distribution, triggering \tool to re-score patients to match the specified distribution.}
    \Description{}
    \label{fig:target-dist}
\end{figure}

\topic{Selecting a target score distribution}
Zooming out from individual patients, the clinician wants to quickly identify the most urgent cases for exploratory purposes. Toward this, they select right-skewed as the target distribution (\Cref{fig:target-dist}), triggering a redistribution of scores to match the specified distribution. This leads \tool to learn and surface updated scoring rules.

\begin{figure}[htp]
    \centering
    \includegraphics[width=0.85\linewidth]{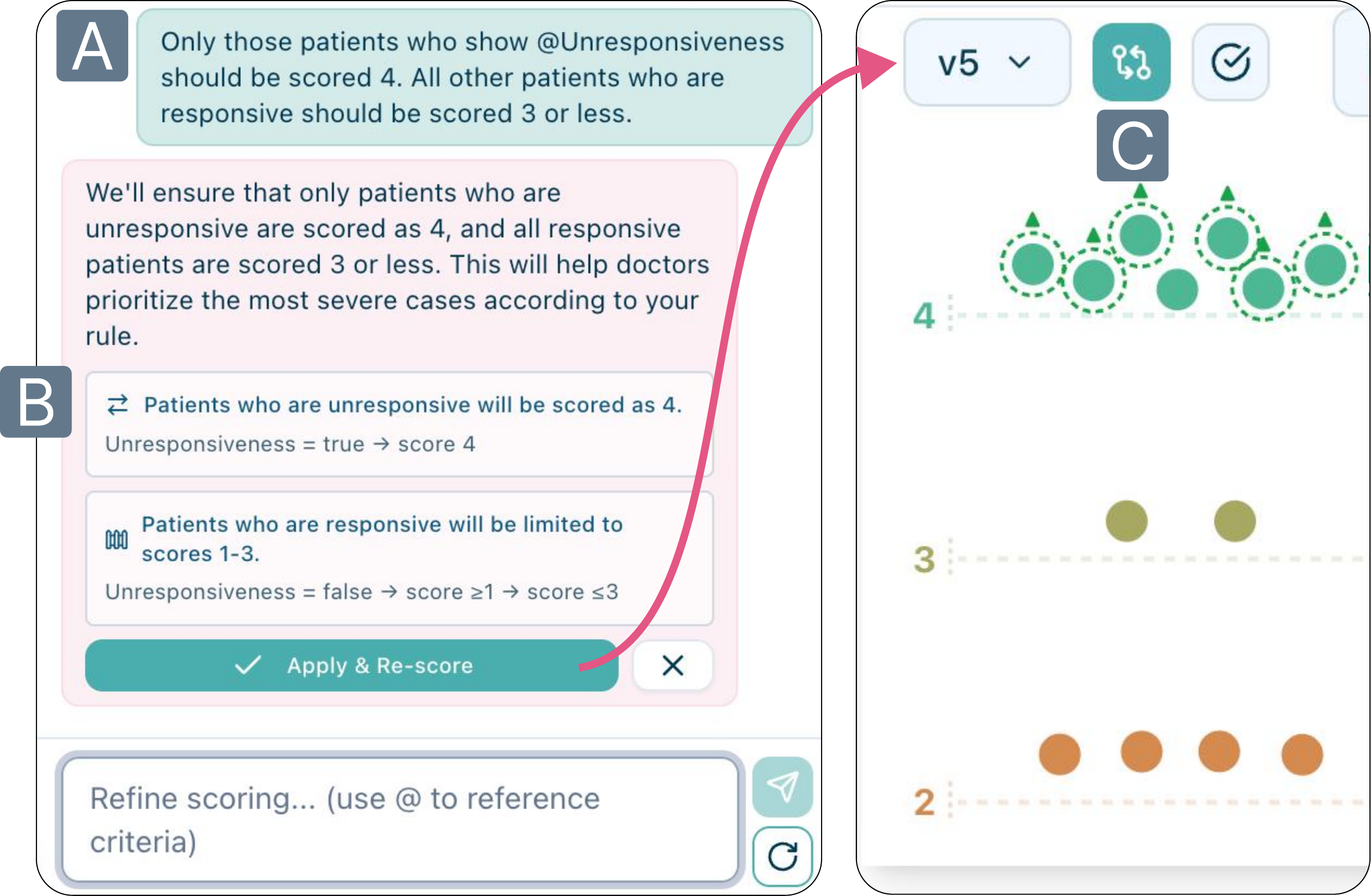}
    \caption{Steering via NL. (A) User provides steering feedback in natural language. (B) \tool{} offers a two-step re-scoring plan that is accepted by the user. (C) User inspects affected scores using the diff view.}
    \Description{}
    \label{fig:nl-to-dm}
\end{figure}

\topic{Chat-based refinement to enforce scoring logic at scale}
After reviewing the most urgent patients, the clinician wants to enforce a specific rule: only patients exhibiting unresponsiveness should receive the highest severity score. Instead of manually composing a scoring rule, the clinician delegates control to the AI chat assistant by describing the rule in natural language (\Cref{fig:nl-to-dm}A). In response, \tool{} first proposes a plan (\Cref{fig:nl-to-dm}B) to move unresponsive patients to score group 4, and re-score all responsive patients between 1 and 3. The clinician reviews and approves this plan (\textbf{C3}), and uses the diff view to inspect affected scores (\Cref{fig:nl-to-dm}C).

\begin{figure}[htp]
    \centering
    \includegraphics[width=\linewidth]{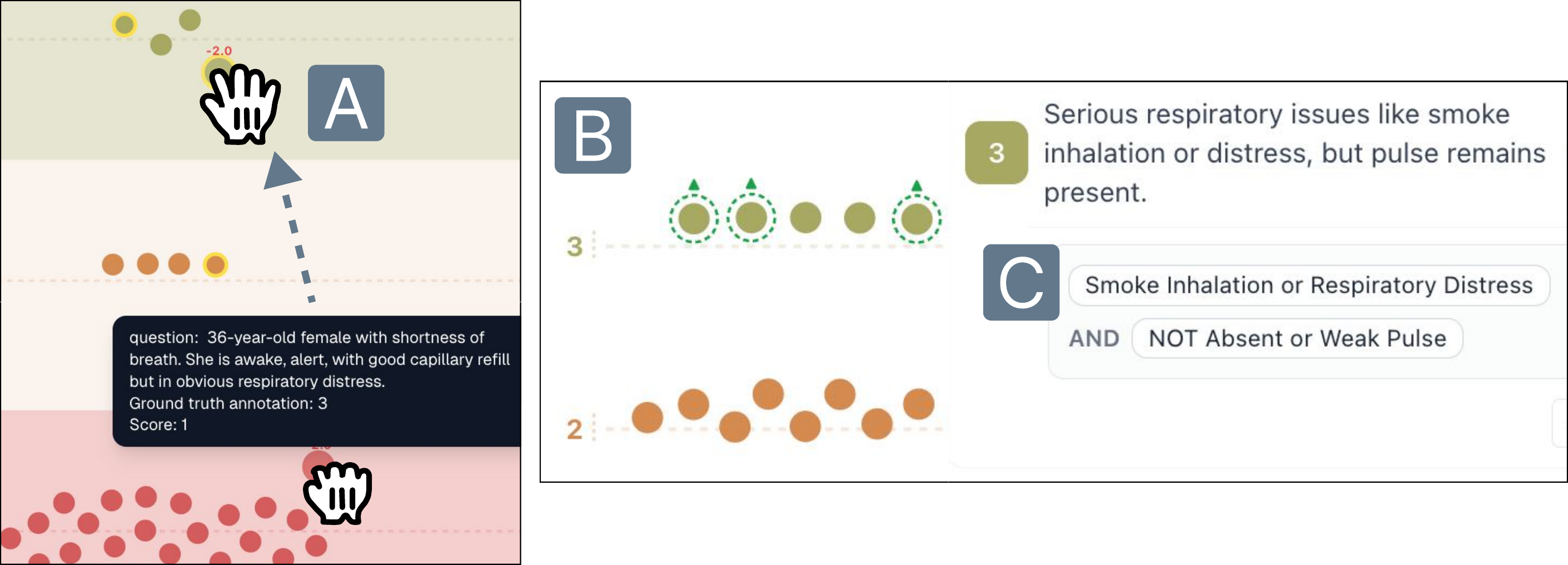}
    \caption{Specifying Examples via Direct Manipulation: (A) User drags and drops input from score 1 to 3. (B) User triggers a re-score that considers the example as guidance. (C) User inspects updated scoring rule to validate their modification.}
    \Description{}
    \label{fig:example-dm}
\end{figure}

\topic{Drag-and-drop to steer scoring rules by example}
When the clinician reviewed updated scoring rules, they found a patient with respiratory distress stranded in score group 1. Rather than formulating a rule, they directly drag-and-drop the patient to score group 3 in the cluster visualization (\Cref{fig:example-dm}A) and trigger re-scoring, causing two additional patients with respiratory distress to be scored higher (\Cref{fig:example-dm}B). The updated rules (\Cref{fig:example-dm}C) confirm that respiratory distress now characterizes score 3---a single example reshapes the scoring rule in a predictable, legible manner (\textbf{C3}).
\section{\tool: System Implementation}

\begin{figure*}[htp]
    \centering
    \includegraphics[width=1\linewidth]{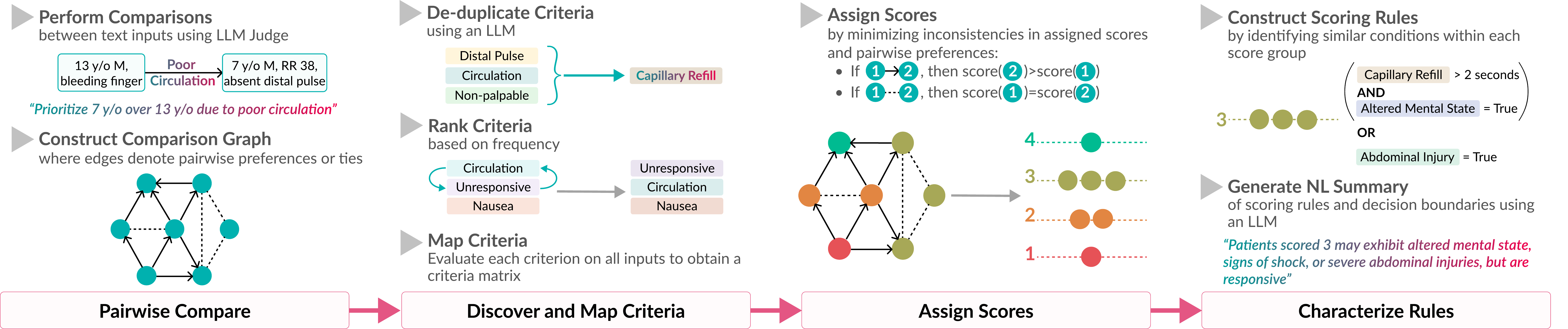}
    \caption{\tool's scoring workflow. \tool{} (1) gathers holistic data understanding through pairwise LLM comparisons and builds a comparison graph. (2) \tool discovers scoring criteria bottom-up from rationales articulated during comparisons, and
    (3) derives consistent score assignments via integer linear programming. (4) Finally, it characterizes scoring rules for each group.}
    \Description{}
    \label{fig:scoring_algorithm}
\end{figure*}

\tool{} is implemented as a web application 
with a Next.js frontend and Python FastAPI backend, using server-sent events (SSE) for streaming progress and LiteLLM to route LLM invocations to OpenAI gpt-4.1. \add{\tool's live deployment and source code are available at \textcolor{blue}{\href{https://github.com/ucbepic/attune}{github.com/ucbepic/attune}}.} The interface provides tri-directional linking between inputs, criteria, and scoring rules, and captures a snapshot after each refinement to support diff views and incremental validation (\textbf{C2}, \textbf{C4}).
\tool's algorithm divides scoring into a one-time initialization phase and an iterative steering loop (Algorithm~\ref{alg:attune}, Figure~\ref{fig:scoring_algorithm}). Given records $R$, a natural language task description $T$, and a score range $[s_{\min}, s_{\max}]$, the initialization phase constructs a comparison graph $G$, discovers scoring criteria $C$, and produces a criteria matrix $M$. The steering loop then assigns scores by solving an integer linear program (ILP), presents the resulting scores, criteria, and rules in the interface, and awaits the user's steering interactions---formalized as constraints fed back into the ILP. The loop repeats until the user is satisfied with the resulting scoring logic. \Cref{tab:notation} in Appendix~\ref{appendix:tech-impl} summarizes notations.

\subsection{Comparative Scoring}
The initialization phase (refer Algorithm \ref{alg:attune}) builds the three structures that underpin consistent scoring and all downstream steering interactions: the comparison graph $G$, the set of scoring criteria, and the criteria matrix $M$.

\begin{algorithm}[htp]
\footnotesize
\SetAlgoLined
\KwIn{records $R$, task description $T$, score range $[s_{\min}, s_{\max}]$}
\BlankLine
\SetKwFunction{FAttune}{\tool}
\SetKwFunction{FPairwise}{Pairwise\_Compare}
\SetKwFunction{FDiscover}{Discover\_Criteria}
\SetKwFunction{FMap}{Map\_Criteria}
\SetKwFunction{FAssign}{Assign\_Scores}
\SetKwFunction{FCharacterize}{Characterize\_Rules}
\SetKwFunction{FPresent}{Present}
\SetKwFunction{FAwait}{Await\_User}
\SetKwFunction{FTranslate}{Translate\_NL\_to\_Actions}
\SetKwProg{Fn}{Function}{:}{}
\Fn{\FAttune{$R, T, s_{\min}, s_{\max}$}}{
    \tcp{Initialization Phase}
    $G \gets$ \FPairwise{$R, T$} \tcp*{comparison graph}
    $C \gets$ \FDiscover{$G$} \tcp*{criteria}
    $M \gets$ \FMap{$C, R$} \tcp*{criteria matrix}
    $\mathit{user\_constraints} \gets \emptyset$

    \tcp{Interactive Steering Loop}
    \While{\textnormal{user is not satisfied}}{
        $\mathit{scores} \gets$ \FAssign{$G, \mathit{user\_constraints}$}\;
        $\mathit{rules} \gets$ \FCharacterize{$\mathit{scores}, M$}\;
        \FPresent{$\mathit{scores}, C, \mathit{rules}$}\;

        $\mathit{actions}, \mathit{nl\_feedback} \gets$ \FAwait{}\;

        \tcp{Translate NL feedback into action(s)}
        \If{$\mathit{nl\_feedback} \neq \emptyset$}{
            $\mathit{actions} \gets \mathit{actions} \cup {}$\FTranslate{$\mathit{nl\_feedback}, R, T, s_{\min}, s_{\max}, M$}\;
        }

        \tcp{Interpret all steering interactions}
        \ForEach{$\mathit{action} \in \mathit{actions}$}{
            \textbf{match} $\mathit{action}$:\\
            \Indp
            $\text{set\_example}(i, s') \rightarrow \mathit{user\_constraints} \gets \mathit{user\_constraints} \cup \{s[i] = s'\}$\;
            $\text{set\_bounds}(i, \ell, u) \rightarrow \mathit{user\_constraints} \gets \mathit{user\_constraints} \cup \{\ell \leq s[i] \leq u\}$\;
            $\text{set\_distribution}(\mathit{targets}) \rightarrow \mathit{user\_constraints} \gets \mathit{user\_constraints} \cup \{\mathit{group\_sizes} \approx \mathit{targets}\}$\;
            $\text{remove\_criterion}(\mathit{name}) \rightarrow \text{drop } \mathit{name} \text{ from } G \text{ and } M$\;
            $\text{add\_criterion}(\mathit{name}, \mathit{def}) \rightarrow M \gets M \cup {}$\FMap{$\{(\mathit{name}, \mathit{def})\}, R$}\;
            $\text{edit\_rule}(\mathit{rule} \to s') \rightarrow \forall \textnormal{ matched } i\colon \mathit{user\_constraints} \gets \mathit{user\_constraints} \cup \{s[i] = s'\}$\;
            \Indm
        }
    }
}
\caption{\tool: Consistent and Steerable Scoring} \label{alg:attune}
\end{algorithm}

\topic{\ttt{Pairwise\_Compare}}
To build a holistic understanding of the data, \tool constructs a comparison graph $G = (V, E)$. The \ttt{Pairwise\_Compare} function takes records $R$ and a task description $T$ as input and produces G, such that each node $v \in V$ corresponds to a record, and each directed edge $e \in E$ connects the losing record to the preferred record. For each record pair (A, B), the LLM returns one of three verdicts---A wins, B wins, or tie---alongside a natural language list of reasons for its verdict. These judgments and their reasons serve as the basis for score assignment and criteria discovery downstream.

Performing all $\binom{|R|}{2}$ comparisons is prohibitively expensive, so \tool builds a sparse graph instead. It first samples a random spanning tree guaranteeing global connectivity, and then augments it with random edges until reaching a target density of $|R|(|R|+1)/4$ edges
(25\% of all possible pairs).

\topic{\ttt{Discover\_Criteria}}
\tool takes the comparison graph $G$ with its per-edge criteria annotations and produces a set of typed scoring criteria (boolean or integer).
The system aggregates all criteria annotations across edges of $G$, and passes them to the LLM to resolve duplicates and organize them to be monotonic, diverse, and typed. The system then ranks the criteria based on their frequency of appearance. These criteria $C$ are presented in \tool's interface as a shared representation of scoring logic (\Cref{fig:interface}D). \add{We note that because \tool performs inductive criteria discovery, it can only surface dimensions actually exhibited by records, and does not guarantee completeness of the criteria set $C$.}

\topic{\ttt{Map\_Criteria}} 
\tool{} now takes the discovered criteria $C$ and records $R$ to produce a criteria matrix $M$. To populate this matrix, it invokes an LLM per criterion and record to obtain a discrete integer or boolean value depending on the criterion type.

\topic{\ttt{Assign\_Scores}} Given the comparison graph $G$ and any accumulated user constraints, \tool{} translates relative pairwise judgments into absolute scores on a fixed scale $[s_{\min}, s_{\max}]$. We formulate score assignment as an integer linear program (ILP).
Formally, for each record $i$, 
define an integer decision variable $s[i] \in [s_{\min}, s_{\max}]$, and for each pairwise judgment indexed by $\text{idx}$, a binary violation variable $v[\text{idx}] \in \{0, 1\}$ that equals 1 when the score assignment contradicts the judgment. The objective minimizes the number of contradicted pairwise judgments:

\begin{equation}
\min \sum_{\text{idx}} w_{\text{judgment}} \cdot v[\text{idx}]
\tag{1}
\end{equation}

\noindent \add{The penalty weight, $w_{\text{judgment}}$, is set to be twice as strong when a tie is violated because, intuitively, a tie asserts both: A is not preferred to B, and B is not preferred A, whereas a win or loss judgment specifies only one ordering. This helps us mitigate unwanted over-differentiation from comparisons.}
We now minimize Equation~1 with respect to constraints from pairwise judgments in the comparison graph. Because LLM judgments may be intransitive---e.g., the LLM may judge $A > B$, $B > C$, but $C > A$---hard constraints would render the ILP infeasible. We therefore encode them as soft constraints using the big-$\mathcal{M}$ method, where $\mathcal{M} = s_{\max} - s_{\min} + 1$:
\begin{align}
\text{If } A \text{ wins:} \quad & s[A] - s[B] \geq 1 - \mathcal{M} \cdot v[\text{idx}] \tag{2a} \\
\text{If } B \text{ wins:} \quad & s[B] - s[A] \geq 1 - \mathcal{M} \cdot v[\text{idx}] \tag{2b} \\
\text{If tied:} \quad & |s[A] - s[B]| \leq \mathcal{M} \cdot v[\text{idx}] \tag{2c}
\end{align}
When a constraint is satisfied ($v[\text{idx}] = 0$), it enforces the judgment: Equation~2a requires the winner's score to be strictly higher; Equation~2c requires tied records to share the same score. When violated ($v[\text{idx}] = 1$), the big-$M$ term relaxes the constraint, allowing the assignment to proceed at the cost of the violation penalty in Equation~1.\footnote{\add{Concretely, when $v[\text{idx}] = 1$, Equation~2a's right-hand side becomes $s_{\min} - s_{\max}$, which is the lower bound of $s[A] - s[B]$ over the score domain and is therefore satisfied by any feasible assignment, including those with $s[A] < s[B]$. The value $\mathcal{M} = s_{\max} - s_{\min} + 1$ is the smallest constant for which this holds.}}
This allows the ILP to find a globally consistent assignment even when individual judgments conflict.

\topic{\ttt{Characterize\_Rules}}
This step takes the score assignments and criteria matrix $M$ to produce human-readable scoring rules for each score group $g$. \tool presents the resulting conjunctive conditions (e.g., ``Capillary Refill $> 2$ AND Altered Mental State $=$ True''), natural language summaries, and decision boundary descriptions in its interface (\Cref{fig:interface}E). \rebecca{For set\_distribution, we should define what $x[i, g]$ means}

\tool{} characterizes rules in three stages. First, it identifies discriminating conditions: each criterion yields one or more testable conditions (e.g., ``Unresponsive $=$ True'' for a boolean criterion; ``Capillary Refill $> 2$'' for an integer criterion). For each condition, the system computes its lift for a given score group $g$---i.e., the ratio of its prevalence within $g$ to prevalence outside $g$. 
Second, \tool{} constructs rules using a greedy set cover strategy: for each score group $g$, it selects the highest-lift condition first, and continues to add conditions until the conjunction still covers at least 50\% of uncovered records in $g$. Once a rule is formed, the records it matches are set aside, and the process repeats to produce additional rules for the same group until all records are covered by a rule. Third, \tool{} invokes an LLM to generate a natural language description for each score group alongside transition phrases that describe how adjacent groups differ (e.g., ``Score 3 patients differ from Score 2 by exhibiting respiratory failure or severe bleeding without loss of consciousness''). Appendix~\ref{appendix:tech-impl} formally outlines the \ttt{Characterize\_Rules} algorithm. 

\subsection{Steering Interactions}

\add{\tool supports refinements to the scoring logic at multiple granularity levels using a second set of constraints that encode user steering actions (Table~\ref{tab:steering}) (\textbf{C2}). Each steering interaction is encoded as a hard constraint, soft constraint, or, as edits to the shared representations (i.e., the comparison graph $G$, scoring criteria $C$, or criteria matrix $M$).}

\topic{Hard constraints} \add{Interactions \ttt{set\_example}, \ttt{set\_bounds}, and \ttt{edit\_rule} (Figures \ref{fig:nl-to-dm}, \ref{fig:example-dm}) directly pin the score variables $s[i]$: \ttt{set\_example} fixes a record $i$ to an exact score ($s[i] = s'$); \ttt{set\_bounds} restricts it to a range ($\ell \leq s[i] \leq u$); and \ttt{edit\_rule} applies the hard constraint $s[i] = s'$ to each record matched by the edited rule. Because these refinements are enforced as hard constraints, they take precedence over pairwise constraints.}

\begin{table}[htp]
\centering
\footnotesize
\caption{User steering actions and downstream constraints or effects.}
\Description{}
\label{tab:steering}
\begin{tabular}{@{}p{0.22\linewidth} p{0.32\linewidth} p{0.32\linewidth}@{}}
\toprule
\textbf{Action} & \textbf{Description \& Invocation} & \textbf{Formalization} \\
\midrule
{set\_example}($i$, $s'$) & Fixes record $i$ to score $s'$ (drag-and-drop/chat). & $s[i] = s'$ (Hard) \\
\addlinespace
{set\_bounds}($i$, $\ell$, $u$) & Constrains record $i$ to a score range (chat). & $\ell \leq s[i] \leq u$ (Hard) \\
\addlinespace
{edit\_rule}($r \to s'$) & Fixes all records matched by rule $r$ to $s'$ (panel/chat). & $s[i] = s' \;\; \forall i \in r$ (Hard) \\
\addlinespace
{set\_distribution}($t_g$) & Targets size $t_g$ for score group $g$ (templates/chat). & $\sum_i x[i,g] \approx t_g$ (Soft) \\
\midrule
{add\_criterion}($n$, $d$) & Adds a new criterion top-down (panel/chat). & Extends $C$ and $M$ \\
\addlinespace
{remove\_criterion}($n$) & Drops criterion from consideration (panel/chat). & Removes from $C$, $M$, and $G$ \\
\bottomrule
\end{tabular}
\end{table}

\topic{Soft constraints} Interaction \ttt{set\_distribution} shapes the overall score distribution (\Cref{fig:target-dist}). For each score
group $g$, we introduce non-negative variables $\text{over}_{g}$ and $\text{under}_{g}$ to capture how far the group size exceeds or falls short of the target group size $t_g$. 
These deviations are added to the objective in Equation~1 as a second
penalty term:

\begin{equation}
\min \underbrace{\sum_{\text{idx}} w_{\text{judgment}} \cdot v[\text{idx}]}_{\text{pairwise consistency}} + \underbrace{ \sum_{g} (\text{over}_g + \text{under}_g)}_{\text{distribution fit}}
\tag{1'}
\end{equation}
\add{The ILP now minimizes the weighted sum of judgment violations and
distribution deviations, treating target distributions as soft constraints that other user actions can override.}

\topic{Edits to shared representations} Rather than adding constraints to the ILP, \ttt{add\_criterion} and 
\ttt{remove\_criterion} modify the comparison graph $G$, scoring criteria $C$, and criteria matrix $M$. Adding a criterion extends $C$ and $M$ via with an incremental \ttt{Map\_Criteria} invocation for the newly added criterion; while removing a criterion drops the criterion from $C$ and
$M$, along with any edges in $G$ whose annotations referenced it. These
edits reshape the inputs to \tool's scoring workflow.

\topic{Translate NL to Actions}
When users share natural language feedback through the AI Assistant (\Cref{fig:interface}E), \tool{} invokes an LLM to decompose the user's feedback along with any selected context records, into a sequence of steering interactions presented in~\Cref{tab:steering}. For instance,~\Cref{fig:nl-to-dm} demonstrates how the user's feedback to score unresponsive patients a 4 is translated to two actions:
\begin{itemize}[nosep, leftmargin=*]
\item $\forall i$ where $M[i, \text{Unresponsive}] = \texttt{true}$: set\_example($i$, 4) 
\item $\forall i$ where $M[i, \text{Unresponsive}] = \texttt{false}$; set\_bounds($i$, 1, 3).
\end{itemize}
The assistant surfaces this plan for user approval before returning, maintaining human oversight (\textbf{C3}).

At each iteration of the steering loop, the ILP is re-solved with the full, updated set of pairwise and user constraints, producing new score assignments that reflect both the LLM's comparative judgments
and the user's cumulative steering refinements.
\section{Technical Evaluation}
\label{sec:tech-eval}

We evaluate whether \tool's steering interactions are more effective compared to prompt-based scoring when given equivalent user feedback. Across three workloads, we simulate stylized steering interactions---providing the same information (examples, distributions, and rules) to both \tool and a prompt-optimization baseline---and measure which approach better translates feedback into accurate scores. 

\topic{Datasets} We evaluate on 3 datasets with expert-annotated ground truth, spanning three domains and scoring scales (Table~\ref{tab:datasets-eval}). \add{We score all 86 \textsc{Triage} records, and stratified random samples of $N=500$ records each for ASAP and WANDS to preserve each dataset's ground-truth score distribution; all records are drawn from a single essay writing prompt or search query for meaningful comparisons.}

\begin{table}[htp]
\centering
\small
\caption{Datasets used in the technical evaluation.}
\begin{tabular}{llcc}
\toprule
\textbf{Dataset} & \textbf{Task} & \textbf{Range} & \textbf{N} \\
\midrule
\textsc{Triage}~\cite{kirch2024triage} & Score patient injury severity & 1--4 & 86 \\
\textsc{ASAP} Set 1~\cite{mathias2018asap} & Score essay persuasiveness & 1--6 & \del{100}\add{500} \\
\textsc{WANDS}~\cite{chen2022wands} & Score product relevance & 0--2 & \del{100}\add{500} \\
\bottomrule
\end{tabular}
\Description{}
\label{tab:datasets-eval}
\end{table}

\begin{table*}[t]
\centering
\small
\begin{tabular}{ll cccc cccc cccc}
\toprule
& & \multicolumn{4}{c}{\textsc{Triage}} & \multicolumn{4}{c}{\textsc{ASAP}} & \multicolumn{4}{c}{\textsc{WANDS}} \\
\cmidrule(lr){3-6} \cmidrule(lr){7-10} \cmidrule(lr){11-14}
& & Seed & Examples & Distribution & Rules & Seed & Examples & Distribution & Rules & Seed & Examples & Distribution & Rules \\
\midrule
& Baseline & 0.36 & 0.46 & 0.44 & 0.78 & \del{0.08}\add{0.12} & \del{0.32}\add{0.31} & \del{0.44}\add{0.46} & \del{0.58}\add{0.41} & \textbf{\del{0.39}}\add{0.54} & \del{0.33}\add{0.54} & \del{0.32}\add{0.52} & \del{0.53}\add{0.56} \\
& \textsc{Attune} & \textbf{0.60} & \textbf{0.63} & \textbf{0.58} & \textbf{0.86} & \textbf{\del{0.14}\add{0.18}} & \textbf{\del{0.51}\add{0.43}} & \textbf{\del{0.52}\add{0.61}} & \textbf{\del{0.70}\add{0.58}} & \del{0.32}\textbf{\add{0.56}} & \textbf{\del{0.46}\add{0.57}} & \textbf{\del{0.61}\add{0.64}} & \textbf{\del{0.62}\add{0.62}} \\
\bottomrule
\end{tabular}
\caption{Accuracy (exact match) across three workloads and four conditions. Prompt baseline uses optimized prompts; \tool uses comparative scoring with corresponding steering interactions.}
\Description{}
\label{tab:results}
\end{table*}

\topic{Baseline} We use GEPA~\cite{agrawal2025gepa}: a prompt optimization framework that begins with a seed prompt provided by us to score records, and reflects on mis-scored records to propose improved prompts. For each task, we start with a seed prompt and optimize it with GEPA under the same feedback each condition provides to \tool. All seed and optimized prompts are presented in Appendix~\ref{appendix:tech-eval-prompts}.

\topic{Conditions} We compare the accuracy of each approach under four conditions, each providing identical information and feedback. GEPA is provided a budget of ($N\times10$) metric calls for each condition alongside a held-out valset of 25 samples.
\begin{itemize}[nosep, leftmargin=*]
    \item Seed: Using hand-written prompts for the baseline and as \tool's task description in the absence of any steering feedback.  
    \item Examples: Randomly sampled one example per score group using ground truth. The baseline uses these examples as its training set and optimizes the prompt for accuracy. \tool receives the same examples as \texttt{set\_example} actions.
    \item Distribution: Provides the target score distribution using ground truth. GEPA optimizes the prompt to minimize deviation from this distribution, i.e., the same objective \tool uses for distribution fit using the \texttt{set\_distribution} action.
    \item Rules: Provides natural language scoring rubrics and guidelines detailed by each dataset. These replace the seed prompt for the baseline condition. For \tool, the authors manually map the provided rubrics to \tool's derived scoring criteria and present them as \texttt{edit\_rule} actions using the chat assistant.
\end{itemize}

\topic{Results}
Table~\ref{tab:results} reports accuracy across all conditions. We see that \tool \add{consistently} outperforms the baseline\del{in 11 of 12 conditions}. We observe that \tool's steering interactions are more effective than prompt-based refinements. With examples, \tool outperforms the baseline by 3--17 points, showing how providing examples enables propagation of changes to score assignments over the comparison graph. For distributions, \tool yields gains up to 15 points (ASAP). For rules, both approaches benefit from scoring rubrics, but \tool leads by 8--17 points---reflecting the advantage of compiling rules into deterministic constraints rather than relying on LLM rubric adherence, which is a known challenge~\cite{murugadoss2025evaluating, wang2025trustjudge}. \add{Appendix~\ref{appendix:model-variance-exps} evaluates inter-model agreement over pairwise comparisons and resulting score assignments, and overlap in scoring criteria.}
\section{User Study}
\label{sec:findings}

To evaluate how domain experts steer scoring logic using \tool's direct manipulation and chat-based interactions, and whether they develop sufficient trust in the resulting scores for practical use, we conducted a summative study with 8 participants (P1--P8).

\topic{Methodology}
We recruited domain experts in healthcare, education, law, and LLM-based evaluation via snowball sampling~\cite{naderifar2017snowball}. 
All participants consented to recording sessions for transcription and analysis. Each 60-minute remote session comprised a feature walkthrough of \tool{}, followed by observing participants complete a scoring task best aligned with their domain of expertise in 35--40 minutes. Participants were encouraged to think-aloud. \add{All participants were exposed to the same condition and a single task to avoid carry-over effects from learned refinement strategies, and to account for variance in subjective expert opinions.} We concluded each session with closing questions and a post-study questionnaire. 
Full task and participant details provided in Appendix~\ref{appendix:summative-details}. Two authors analyzed transcripts, notes, and screen recordings through reflexive thematic analysis with open and axial coding~\cite{braun2006using, braun2019reflecting}, followed by a second iteration to consolidate themes.

\topic{Overview of usage} Across all sessions, participants followed a consistent workflow: they began by randomly inspecting inputs across score groups, followed by inspecting and refining scoring criteria, and then steering scoring logic through a combination of direct edits and chat-based refinements. Participants collectively annotated 3--8 inputs each to build their validation set, added 1--2 new criteria (P1, P3, P4, P5, P7), and deleted 1--3 criteria (P2, P4, P5, P6, P8).
Notably, criteria validation was a one-time activity---participants rarely revisited the scoring criteria to make changes after an initial inspection---while steering interactions occupied
the bulk of subsequent actions. All participants inspected scoring rules at least once and issued 1--3 chat-based refinements, while four participants (P1, P3, P5, P6) used the target distribution widget. 

\topic{Overview of feedback} 
On 1--5 Likert scale questions, participants rated the quality of scoring favorably ($\mu=4.0$, $\sigma=0.8$) and reported feeling in control of scoring decisions ($\mu=4.0$, $\sigma=1.0$). Scoring criteria and rules were the most highly rated features, followed by annotating samples, providing examples, and specifying target score distributions (see Appendix~\ref{appendix:summative-details} for post-study responses).

\subsection{Users Establish Common Ground Before Steering Scores}
Participants began their workflow by {\em ``spot checking''} inputs and scores, 
matching their own assessments against the system's, 
annotating cases where they disagreed, 
and inspecting criteria for coverage and correctness. 
These early interactions---rapid inspections, 
building a validation set through annotations, and
criteria validation---anchored all subsequent refinements.

\subsubsection{Users build validation sets and confirm criteria coverage through spot checks}
Participants' first instinct was to inspect inputs across the score range using the cluster view and data table to navigate between scores and individual records. While inspecting records, participants compared their own assessment of each input against the system's assigned score. When they disagreed, they organically annotated the record with their own score, accumulating a small validation set (P1--P5, P7, P8). P1 observed that {\em ``annotating even 4--5 patients is helpful to measure alignment across re-score attempts,''} and 
P2 similarly valued having a growing validation set to track progress over successive refinements. 
During spot checks, participants also triangulated with scoring criteria: as they read individual inputs, they cross-referenced features they noticed against the derived criteria to see if \tool captured and matched their understanding (P2--P5). P3 systematically read 1--2 essays per score group, checking what each criterion evaluated to, verifying that features they liked or disliked were captured by the criteria. 
P5 adopted a complementary strategy, spot checking essays belonging to each rule. 
A small number of criteria additions occurred at this stage (P1, P3, P4, P5)---P1 added ``Age'' to complement the existing ``Blood Loss'' criterion after noticing that blood loss is more severe for pediatric patients than adults, 
and P3 added criteria for opening hooks and grammatical errors after encountering essays where these features seemed to matter but were not yet represented.

\subsubsection{Bottom-up scoring criteria earn user trust quickly}
Participants overwhelmingly agreed with the criteria \tool{} derived, finding them to be exhaustive and necessary for their scoring task (P1--P6).
P5 appreciated seeing the criteria upfront: {\em ``I like that the criteria are curated based on actual content of the student essays. When I go in without a rubric, or with one
that was created upfront and handed down to me, it is easy to be harsh for
the first couple essays.''} P3 echoed this, noting surprise at how well
the system handled diversity: {\em ``I wasn't expecting to see such a wide
range of essay quality in the data. It is great that the system still made sense
of this diversity and scored this essay 1.''} 
When participants introduced new criteria (P2, P5, P7) and saw that they evaluated to \ttt{False} across all inputs, it quickly reassured them that \tool's scoring criteria were exhaustive.

We seldom observed criteria deletions during this phase. 
Participants removed criteria that were either overly specific, 
redundant, 
or overlapping (non-independent). 
For instance, P2 deleted ``Absence of Broad Indemnity Clauses'' because it applied to only one lease agreement and was speculative. 
P5 deleted two criteria promoting overly specific examples to be part of student essays.
P4 attempted to distill the criteria set to independent variables by deleting ``Addresses Both Sides'' on seeing ``Addresses Positives'' and ``Addresses Negatives''.    
Once validated, participants treated scoring criteria as grounding artifacts for subsequent refinements.

\subsection{Users Co-Evolve Scoring Logic with \tool}
\tool{}'s steering interactions served two key refinement intentions: {\em (i)} adjusting the overall distribution and placement of scores, and {\em (ii)} reshaping how criteria composed into rules at a granular level. Direct manipulation interactions gave users immediate control over the former, while participants used a combination of direct manipulation and delegation to \tool{}'s chat assistant for complex rule wrangling.

\subsubsection{Scoring criteria and rules let users catch misalignments at-a-glance}
Because scoring criteria and rules were linked to inputs, participants could quickly identify where the system's logic diverged from their expectations. The most common pattern was clicking on a criterion to see how inputs matching that criterion were distributed across score groups (P1--P8). P5 clicked on ``Address Negative Effects'' and immediately noticed that these essays were scored 7, while essays that ``Address Positive Effects'' were scored 8: {\em ``I am curious about the difference between 7 and 8. Currently the LLM seems biased towards essays that highlight positives of computer usage.''} 
P1 systematically clicked on each clinical criterion to check whether patients with respiratory distress and altered consciousness were landing at appropriate severity levels.

Participants also caught misalignments by working outward from inputs they already understood. They used their annotated samples to inspect the linked rules and most salient criteria, ensuring that the system gave the same score to other similar records (P3, P6, P7). 
P6 used this pattern extensively after making refinements to confirm that records similar to the ones they had anchored landed in the same group.
A third pattern was walking through scoring rules systematically. Participants read the conjunctive conditions for each score group and checked whether they matched expectations. P1 inspected rules four times across the session, verifying clinical guidelines were reflected.

\subsubsection{Users show and tell: steering interactions span direct edits, chat, and combinations of both}
Once misalignments were identified, participants blended direct manipulation with chat-driven delegation to steer refinements.
Participants {\em showed} the system what they wanted through distribution controls and drag-and-drop examples. P3 targeted a right-skewed distribution for essay scores; P6 adjusted the distribution twice to reshape recommendations; P2 sought a right-skew for lease contracts to separate acceptable from problematic agreements. At a granular level, P6 dragged a movie recommendation 
to score group 8 to recalibrate scores, and P3 dragged the best essay to 9 to anchor the top of the scale.

Participants {\em told} the system what to change when edits involved groups of inputs or complex rule conditions. 
P8 verbalized this as {\em ``It's easier to tell something to the chat assistant when I want the rules to be different.''}
P4 wrote {\em ``Give less importance to relevant example count, and more importance to persuasive reasoning,''} to the chat assistant. P1 used chat to encode a clinical guideline that expectant patients should receive the lowest priority. Participants shared underspecified intents, asking the chat assistant to {\em ``rethink''} how something should be scored or using qualifying language like {\em ``somewhat higher,''} and then reacted to the proposed plan.

Several participants also combined showing and telling (P2, P3, P6). Such refinements involved selecting context inputs, referencing a criterion, and then writing a chat message. P2 mixed examples, criteria, and target distribution in a single message---{\em ``This lease agreement has so many penalties (paying penalties of 100K for each calendar day). Score all such agreements 1''}---and accepted the resulting rule (If Absence of Severe Penalties = False $\rightarrow$ Score = 1). P6 dragged a movie to 8 and immediately followed up: {\em ``find me movies similar to Spiderman and score them all 8.''}  P3 chained together an example, a chat message to specify scoring bounds, and a distribution target in the same re-scoring attempt.

\subsubsection{Users ground trust in predictable edits}
The chat assistant's plan-based automation was central to participant trust. P2 found that {\em ``the plan made changes feel predictable rather than completely random,''} and P3 noted: {\em ``It was transparent about what it will do, and felt more based in actual context. Based on my prelim checks it adhered to the criteria I mentioned unlike ChatGPT.''} Because participants could anticipate what would change before it happened, validation was lightweight: they inspected scoring rules to understand if \tool was able to generalize well (P1, P3, P5, P6, P7), and re-checked annotated samples (P1, P3, P4, P6). This predictability may have led participants to compose sequences of multiple steering interactions within a single re-scoring attempt rather than making isolated, cautious edits (P3, P6).

\section{Discussion}
\label{sec:discussion}
We now discuss broader applications and appropriate use of \tool, potential overreliance, additional steering interactions based on experts' usage, and other future directions.

\topic{Scope and appropriate use} Several participants envisioned broader applications of \tool{}. P3 described how \tool{} could expose multiple {\em ``scoring functions''} across teaching assistants and help reconcile differences between them. Future work could extend \tool for collaborative calibration, which also connects to prior work in debiasing crowdsourced annotations~\cite{zhuang2015debiasing}. 
P1 noted that \tool could serve as a pedagogical tool to train new medical professionals to reason through clinical prioritization. 
Participants also brought up new application domains. P6 wanted to use \tool to filter relevant job postings, and find rental apartments. P7 proposed exploring the use of \tool{} for police routing and prioritization in emergencies for dynamic matching of responder experiences to specific situations.
These suggestions point toward a broader design space in which \tool{}'s scoring surfaces serve not only as individual steering tools but as shared artifacts for coordination, training, and quality assurance.

\add{That said, several participants reflected on the need for human oversight in sensitive and high-stakes domains such as healthcare, education, and recruitment (P1, P5, P7). Prior work in medicine and education has shown that AI can augment human assessments by serving as a second opinion~\cite{beede2020human, xiao2025human}. \tool similarly makes LLM-powered scoring logic legible and reliably editable, and introduces positive friction as users vet identified criteria and rules (§7.1.1, §7.2.3). Consistent with Schön’s reflection-in-action theory~\cite{schon2017reflective}, this slowdown may help mitigate overreliance, encouraging careful consideration of the rules and incremental testing. We do not propose using \tool to automate scoring and decision making in the absence of human oversight.}

\add{Participants also reflected on the appropriate use of target distributions. P3 and P5 found that enforcing a distribution is often desirable when grading student responses on a curve, or when screening resumes for fixed slots. However, P1, P2 and P7 found that in sensitive domains where the distribution must be learned bottom-up (e.g., clinical triage), this feature can be {\em ``objectionable,''} but still be useful as an {\em ``exploratory probe''} for epistemic reasoning.}

\topic{Biases and lack of faithfulness in LLM rationalizations}
\add{\tool derives criteria and rules from rationales LLMs articulate during pairwise comparisons. However, these are post-hoc rationalizations, and recovering the true cause of an LLM judgment remains difficult. We therefore treat criteria and rules as useful patterns and as handles for steering, and not causal explanations. Additionally, we note that reducing rich, qualitative signals to scores, and presenting scoring criteria and rules as artifacts of scoring decisions risks a false sense of objectivity, carrying known risks of algorithmic decision-making such as automation bias, abstraction traps, and reification of systemic biases~\cite{selbst2018fairness}. Future work could (i) explore mechanistic explanations for greater reliability (e.g., using sparse autoencoders~\cite{movva2025sparse}); (ii) measure trust and overreliance on such systems; and (iii) design interactions that resist overreliance by encouraging counterfactual reasoning of criteria and rules.}

\topic{Extending the design space of steering interactions} Our user study validated the effectiveness of \tool's steering interactions, and also revealed where they fell short. Participants frequently wanted to edit criteria definitions rather than simply accepting or deleting them: P5 spent time reverse-engineering what integer criteria like ``Number of Relevant Examples'' actually meant to the LLM, and several participants tried deleting and re-adding criteria with new definitions. P1 and P8 wanted example-driven criteria redefinitions where they could provide instances and let the LLM infer the definition to lower challenges with crisp articulation. Participants sometimes went back-and-forth with the chat assistant to fix proposed re-scoring plans. P4 wanted to directly edit the proposed plan. Participants also wanted \tool to proactively highlight consequences of drag-and-drop steering interactions ({\em e.g.}, {\em ``I see you moved this to 4---would you like to move all similar inputs to 4?''}). These unmet needs point to opportunities for richer affordances to edit criteria semantics, manipulate proposed plans, and receive proactive suggestions.

\topic{Scoring rule compositions are domain- and context-dependent} 
Our findings surfaced divergences in how experts expected criteria to compose into scoring rules. 
P1, an emergency surgeon, wanted deeply
nested conjunctions with adjustable depth, because the same criterion can carry different clinical significance across scenarios, for example, {\em ``altered
consciousness matters more in alcohol overdose than in trauma, and high blood pressure must be weighed more heavily for a pregnant patient.''} 
P3, a teaching assistant, preferred an additive model ({\em e.g.}, +2 for reflecting on personal experience,
$-$1 for each grammatical error), and found conjunctive rules unnatural for essay grading.
P5 compared \tool to Gradescope, noting that rule tracking is easier when
student responses vary in limited ways, but {\em ``here, it's much more
qualitative.''} 
These differences suggest that rule composition is a domain-dependent decision.
While experts across domains liked the NL representations of score groups and decision boundaries,
future work should explore parameterizable rule templates that explore and accommodate varied types of compositions, and investigate how the choice of rule composition affects users' ability to inspect and trust scoring logic.

\topic{Limitations}
We gathered qualitative findings by observing domain experts in four fields, limiting the generalizability of our findings. Longitudinal engagements with experts from diverse domains could reveal how scoring rules evolve as they encounter streaming data with new characteristics. Future work could also investigate strategies for sampling representative records, adaptive sampling of pairwise comparisons for cost and scalability, and empirically evaluate the generalizability of scoring criteria and rules at scale.
\section{Conclusion}

We presented \tool, a mixed-initiative system for steerable LLM-powered scoring. Informed by a formative study, \tool{} derives scoring criteria and rules bottom-up through pairwise comparisons, and compiles user steering interactions—examples, criteria edits, rule modifications, and distributional targets—into deterministic ILP constraints that guide re-scoring. Our technical evaluation showed improved accuracy over prompt-based scoring across three workloads, and a user study with domain experts confirmed that participants trusted bottom-up criteria quickly, composed steering interactions fluidly, and grounded confidence in the predictability of refinements.

\balance

\begin{acks}
\add{We are grateful to Sewon Min, Joseph M. Hellerstein, Austin Z. Henley, Emma Pierson, Rishabh Tiwari, and Lakshya A Agrawal for their valuable discussions and feedback on our prototypes. We thank our participants for their enthusiastic engagement and feedback. We acknowledge support from grants DGE-2243822, IIS-2129008, IIS-1940759, and IIS-1940757 awarded by the National Science Foundation, funds from the State of California, funds from the Alfred P. Sloan Foundation, as well as EPIC lab sponsors: Adobe, Google, G-Research, Jane Street, Microsoft, PromptQL, Sigma Computing, Snowflake, and Bridgewater. Compute credits were provided by Azure, Modal, NSF (via NAIRR), and OpenAI.}
\end{acks}

\bibliographystyle{ACM-Reference-Format}
\bibliography{references.bib}

\clearpage

\appendix
\section{Design Probe: Implementation Details}
\label{appendix:probe}
\rebecca{Consider deleting this first paragraph because it duplicates information in the System Implementation section}
The design probe used in our formative study was implemented as a full-stack web application. The frontend was built with Next.js and React. The backend was implemented in Python using FastAPI, which handled LLM orchestration. All LLM invocations were routed to OpenAI \texttt{gpt-4.1} via LiteLLM. Server-sent events (SSE) were used to stream scoring progress to the frontend in real time.
The probe captured a version snapshot after each refinement iteration, storing the updated scoring prompt and assigned scores. These snapshots were accessible via a version history dropdown. Selecting a prior version displayed the scores from that iteration. The diff view supported lightweight validation by letting participants identify impacted records at-a-glance.
To score records, the probe used the following batched scoring prompt:

\tealbox{
\textbf{Prompt: Batched Scoring}

Score each of the following records based on the task description.

\medskip
Task: \texttt{\{task\}}

Score range: \texttt{\{min\_score\}} (lowest) to \texttt{\{max\_score\}} (highest)

Records to score:
\texttt{\{records\}}

\medskip
For each item, provide a score as an integer between \texttt{\{min\_score\}} and \texttt{\{max\_score\}} inclusive.
Respond in the following JSON format with one object per item, in the same order as presented:

\noindent \texttt{"scores": \{"id1": <score1>, "id2": <score2>, ...\}}
}

\noindent To refine scoring, users could edit the task description or provide natural language chat feedback (optionally selecting any records as context). If provided chat feedback, the probe would first invoke an LLM to generate an updated task description:

\tealbox{
\textbf{Prompt: Task Description Rewriting}

You are a prompt engineer. The user has a scoring task and wants to refine how records are scored.

\medskip
Current task description:
\texttt{\{current\_prompt\}}

Score range: \texttt{\{min\_score\}} to \texttt{\{max\_score\}}

Previous refinements:
\texttt{\{chat\_history\}}

Selected context: \texttt{\{context\}}

Latest refinement request:
\texttt{\{message\}}

\medskip
Based on the latest refinement request, produce an updated scoring task description that incorporates the user's feedback. The updated description should be a complete, self-contained task description. Keep it clear and specific. If the user referenced specific items as context above, use them as concrete examples or anchors in the updated prompt. Also produce a brief explanation changes.

\medskip
Respond in the following JSON format:

\noindent 
\texttt{\{"task\_description": "...", "explanation": "..."\}}}

\noindent The \texttt{\{context\}} field was populated only when the user had selected one or more records before issuing their refinement. When no records were selected, this section was omitted. The \texttt{\{chat\_history\}} field accumulated a chronological list of all prior chat messages, providing the model with a summary of the user's evolving intent. 
On obtaining the updated task description, the probe re-scored all records using the batched scoring prompt. 

\section{Formative Study: Supplementary Details}
\label{appendix:formative}

\subsubsection*{Participant Backgrounds}
Formative study participants were recruited based on their expertise in using LLMs as judges or evaluators in enterprise pipelines, allowing us to study breakdowns and strategies that arise from informed use rather than novice prompting difficulties. 
Participants included graduate students, research engineers, and scientists from academia and industry. Their expertise spanned evaluating RAG pipelines, assessing AI agent traces, building and evaluating, evolution of automation agents, and model finetuning at technology companies and research groups (\Cref{tab:formative-participants}).

\begin{table}[htp]
\caption{Formative study participant backgrounds and datasets.}
\label{tab:formative-participants}
\footnotesize
\begin{tabular}{llll}
\toprule
\textbf{ID} & \textbf{Role/Experience} & \textbf{Dataset} \\
\midrule
F1 & AI safety and governance & Candidate Resumes \\
F2 & AI for education & Essays (ASAP Set 1) \\
F3 & RAG accuracy benchmarking & \textsc{Wands} \\
F4 & Agent trajectory analysis & \textsc{Triage} \\
F5 & Enterprise AI-product evaluation & Essays (ASAP Set 1) \\
F6 & RAG accuracy benchmarking & Candidate Resumes \\
F7 & LLM and agent benchmarking & \textsc{Triage} \\
F8 & LLM Finetuning & \textsc{Wands} \\
F9 & LLM agent evolution & Essays (ASAP Set 1) \\
F10 & Enterprise AI-product evaluation & \textsc{Triage} \\
F11 & AI for education & Candidate Resumes \\
F12 & Enterprise AI-product evaluation & \textsc{Wands} \\
\bottomrule
\end{tabular}
\end{table}


\subsubsection*{Study Tasks}
Tasks varied along two dimensions: the level of expertise required (generic vs. domain-specific) and the evaluation target (accuracy-based vs. subjective) (\Cref{table:probe-tasks}). For accuracy-based tasks, expert-annotated ground-truth labels were available, and the probe displayed accuracy and mean absolute error (MAE) metrics in the interface header. For subjective tasks, no ground-truth labels were available; instead, evaluation relied on participants' satisfaction with the generated scores. Each task was completed by three participants, allowing us to examine how strategies vary across tasks differing in subjectivity and domain knowledge requirements.

\begin{table}[h]
\centering
\caption{Formative study datasets and tasks.}
\label{table:probe-tasks}
\footnotesize
\begin{tabular}{p{2.6cm} p{5cm}}
\toprule
\textbf{Dataset} & \textbf{Task} \\
\midrule
\multicolumn{2}{l}{\textbf{Objective} \textit{(ground-truth available)}} \\
\midrule
\textsc{Triage}~\cite{kirch2024triage} & Score severity of patient injuries on a scale of 1--4 to prioritize treatment. \\
Essays (ASAP Set 1)~\cite{mathias2018asap} & Score essays that argue whether computers have positive or negative effects based on persuasiveness on a scale of 1--6. \\
\midrule
\multicolumn{2}{l}{\textbf{Subjective} \textit{(ground-truth not available)}} \\
\midrule
\textsc{Wands}~\cite{chen2022wands} & Score monitor stands' product relevance based on personal selection criteria. \\
Candidate Resumes~\cite{bracko2023resume} & Score candidate resumes based on their  relevance for a senior data engineer role. \\
\bottomrule
\end{tabular}
\end{table}

\subsubsection*{Analysis}
We analyzed transcripts supplemented with notes documenting participant actions. Three authors performed reflexive thematic analysis through open coding of the transcripts, notes, and screen recordings, followed by identifying axial codes~\cite{braun2006using, braun2019reflecting}. 
We coded along two dimensions: (1) validation and inspection patterns, as well as observations that triggered refinements, and (2) the types of refinement feedback participants provided to steer scoring.
\section{Implementation: Supplementary Details}
\label{appendix:tech-impl}

Here we provide a summary of notation used in \S5 (\Cref{tab:notation}), prompt templates used to de-duplicate scoring criteria and translate natural language feedback to steering actions, and the pseudocode for the \ttt{Characterize\_Rules} step (Algorithm~\ref{alg:characterize}). 


\begin{table}[ht]
\centering
\caption{Summary of notation.}
\label{tab:notation}
\footnotesize
\begin{tabular}{@{} l p{6cm} @{}}
\toprule
\textbf{Symbol} & \textbf{Definition} \\
\midrule
$R = \{r_1, \ldots, r_n\}$ & Set of records to be scored \\
$T$ & Natural language task description \\
$[s_{\min}, s_{\max}]$ & Integer score range \\
$s[i]$ & Score assigned to record $i$ \\
$G = (V, E)$ & Comparison graph; $V = R$, edges carry pairwise judgments\\
$C = \{c_1, \ldots, c_k\}$ & Set of scoring criteria \\
$M$ & Criteria matrix; $M_{ij}$ is the value of criterion $c_j$ on record $r_i$ \\
$v[\text{idx}] \in \{0,1\}$ & Violation indicator for pairwise judgment idx \\
$\mathcal{M}$ & Big-$M$ constant ($ s_{\max} - s_{\min} + 1$) \\
$w_{\text{judgment}}$ & Violation weight ($w_{\text{tie}} = 2$; $w_{\text{win}} = w_{\text{loss}} = 1$) \\
$t_g$ & Target size for score group $g$\\
$\text{over}_g, \text{under}_g$ & Slack variables for distribution fit \\
$\operatorname{lift}(p, g)$ & Ratio of a predicate $p$'s prevalence within score group $g$ to its prevalence outside $g$ \\
\bottomrule
\end{tabular}
\end{table}

\begin{algorithm}[ht]
\footnotesize
\SetAlgoLined
\DontPrintSemicolon
\KwIn{Scores, criteria matrix $M$}
\KwOut{Rule set $\mathcal{R}$}
\BlankLine
\tcp{Stage 1: Identify discriminating conditions}
\ForEach{score group $g$}{
  \ForEach{criterion $c_j \in C$}{
    Generate candidate conditions from $c_j$ (e.g., $c_j = \texttt{True}$ for boolean; $c_j > v$ for integer)\;
    Compute lift of each condition for group $g$\;
  }
  Rank conditions by lift\;
}
\BlankLine
\tcp{Stage 2: Construct rules via greedy set cover}
\ForEach{score group $g$}{
  $U \gets$ all records in group $g$ \tcp*{uncovered records}
  \While{$U \neq \emptyset$}{
    Initialize rule $\phi$ with the highest-lift condition\;
    \While{conjunction covers $\geq 50\%$ of $U$}{
      Extend $\phi$ with the next highest-lift condition\;
    }
    Add $(\phi \rightarrow g)$ to $\mathcal{R}$\;
    Remove records matched by $\phi$ from $U$\;
  }
}
\Return $\mathcal{R}$\;
\caption{\textsc{Characterize\_Rules}}
\label{alg:characterize}
\end{algorithm}

\tealbox{
\textbf{Prompt: Criteria De-duplication}

\medskip
You are analyzing scoring criteria for the following task:

Task: \texttt{\{task\_description\}}

\medskip
Below is a list scoring criteria, which may be overlapping, or synonymous. You must:
\begin{itemize}[nosep, leftmargin=*]
    \item Group synonymous or overlapping criteria into canonical criteria (2-5 words, Title Case).
    \item List which raw criterion map to each canonical criterion.
\end{itemize}
Any overlapping or duplicate criterion must appear in exactly one de-duplicated criterion's \ttt{overlapping\_criteria} list.

\medskip
Scoring criteria: \texttt{\{scoring\_criteria\}}

\medskip
Output format:

\noindent \texttt{
\{deduplicated\_criterion\_name: overlapping\_criteria=[...], ...\}
}
}


\tealbox{
\textbf{Prompt: Translate NL to Actions}

You are an assistant that interprets user feedback to produce a concrete plan of actions. You may emit a sequence of one or more actions that best address the user's feedback.

\medskip
Scoring task: \texttt{\{task\_description\}}

Score range: \texttt{\{min\_score\}} to \texttt{\{max\_score\}}

Scoring criteria: \texttt{\{scoring\_criteria\}}

Selected context: \texttt{\{context\}}

User feedback:
\textbf{``\texttt{\{message\}}''}

\medskip
Possible actions to emit:
\begin{enumerate}[leftmargin=*]
    \item \texttt{set\_example}: When the user says individual records or records matching a criterion condition should BE a specific score.
    \item \texttt{set\_bounds}: When the user wants to constrain individual records or records matching a criterion condition to a score range (min/max), and not an exact score.
    \item \texttt{set\_distribution}: When the user wants to change the overall shape of scores (e.g. ``bell curve'', ``spread evenly''). Template must be one of: uniform, gaussian, left-skew, right-skew.
    \item \texttt{add\_criterion}: When the user references a concept that does NOT exist as a criterion yet. Do NOT add criteria that already exist.
    \item \texttt{remove\_criterion}: When the user wants a crtierion to stop affecting their task.  
\end{enumerate}

Ordering Guideline: 
Always emit \texttt{add\_criterion} actions BEFORE \texttt{set\_example} or \texttt{set\_bounds} actions that depend on them.
}
\section{Scoring Prompts for Technical Evaluation}
\label{appendix:tech-eval-prompts}

Below, we present the seed and GEPA-optimized prompts used by the prompt baseline for each condition in our technical evaluation (\S6), along with the rules-based chat feedback provided to \tool{}, organized by dataset.

\topic{\textsc{Triage}}

\tealbox{
\textbf{Prompt: Seed}

Score patient injury severity on a scale of 1 to 4 based on first responder notes, with 1 being least severe and 4 being most severe.
}

\tealbox{
\textbf{GEPA-Optimized Prompt: Examples}

You are tasked with scoring the severity of a patient's injury based on first responder notes. The notes are provided in free text and will describe the patient's age, clinical signs, symptoms, and any apparent injuries or conditions. Use the following injury severity scale:

\begin{itemize}[nosep, leftmargin=*]
  \item \textbf{1 -- Minor injury:} No significant impairment or threat to life. Examples: superficial wounds, mild symptoms, stable vital signs, no neurological or vascular compromise.
  \item \textbf{2 -- Moderate injury:} Not life-threatening, but more than minor. Examples: notable soft tissue injuries without neurovascular compromise, full range of motion, sensation intact, no active bleeding.
  \item \textbf{3 -- Serious injury:} Potentially or actually life-threatening, or involves significant compromise, but not immediately requiring resuscitation. Examples: moderate vital sign abnormalities, significant pain or injury, evidence of possible internal injury without shock, burns covering non-critical areas.
  \item \textbf{4 -- Severe/Critical injury:} Life-threatening or requires immediate intervention. Examples: severe vital sign abnormalities (e.g., tachypnea >40, weak pulse, altered mental status), major trauma, signs of shock, severe neurological impairment, compromised perfusion, extensive burns, multiple injuries with instability.
\end{itemize}


If vital sign abnormalities or multiple serious injuries are described, score higher (3 or 4). If injuries are limited, stable, and without complications, score lower (1 or 2). 
}

\tealbox{
\textbf{GEPA-Optimized Prompt: Distribution}

You are tasked with assigning an integer score to items describing patient injuries, as detailed in first responder notes. The scoring is performed on a scale from 1 to 4, representing increasing severity of injury.


\textbf{Calibration:} When scoring across a batch of items (e.g., 100 cases), your scores should approximate the following distribution:
\begin{itemize}[nosep, leftmargin=*]
  \item Score 1 to about 28\% of cases (minor injury/few complications)
  \item Score 2 to about 24\% of cases (moderate, not critical)
  \item Score 3 to about 38\% of cases (serious, non-critical)
  \item Score 4 to about 10\% of cases (life- or limb-threatening or requiring immediate intervention)
\end{itemize}

To maximize accuracy and calibration:
\begin{itemize}[nosep, leftmargin=*]
  \item Carefully evaluate the description for signs of critical injury (vital sign abnormalities, altered mental status, uncontrolled bleeding, complete loss of function, or obvious life-threatening mechanisms). Assign score 4 if any of these are present.
  \item Assign score 3 for considerable mechanisms of injury or wounds that are severe but not demonstrating abnormal vital signs or impaired consciousness; e.g., penetrating wounds with intact pulses and sensation, crushed limbs with preserved perfusion, etc.
  \item Assign score 2 for injuries worse than minor but lacking severe symptoms, e.g., moderate pain, limited injury, controlled bleeding, and no major functional limitations.
  \item Assign score 1 only for patients meeting minor criteria: alert, able to walk, minimal pain or injury, no disturbance of vital signs, no threat to life or limb.
\end{itemize}
}

\tealbox{
\textbf{GEPA-Optimized Prompt: Rules}

You are given a medical triage scoring task. Your goal is to assign a discrete integer score to a described patient's condition using a specified rubric from 1 to 4.

\textbf{Detailed Scoring Rubric} (use this rubric precisely):
\begin{itemize}[nosep, leftmargin=*]
  \item \textbf{Score 1:} Patient has relatively minor injuries. Their status is stable, and deterioration is unlikely over days.
  \item \textbf{Score 2:} Patient has serious and potentially life-threatening injuries, but their status is unlikely to deteriorate significantly over several hours.
  \item \textbf{Score 3:} Patient requires medical attention within 60 minutes or less for survival, due to compromise of airway, breathing, or circulation.
  \item \textbf{Score 4:} Patient is unlikely to survive given the severity of their injuries.
\end{itemize}

\textbf{Domain-specific considerations:}
\begin{itemize}[nosep, leftmargin=*]
  \item ``Compromise to airway, breathing, or circulation'' means any findings or hints in the scenario that the patient cannot breathe appropriately, has very poor or absent pulses, is in shock, or has significant altered mental status due to injury.
  \item Assess vital signs (e.g., abnormal respiratory rate, level of consciousness), circulation (e.g., distal pulses, presence of active bleeding), and mobility (e.g., can ambulate after injury?) as presented.
  \item Severe mechanism of injury with normal vital signs and findings (e.g., impaled objects without bleeding or pulse loss) may not automatically raise the severity score unless there's a plausible risk of rapid deterioration apparent in the description.
  \item Non-ambulatory status, low respiratory rate (especially in children), or only responding to painful stimuli indicate higher severity.
  \item The decision should only be based on the information provided.
\end{itemize}

}

\tealbox{
\textbf{\tool: Rules}

\begin{itemize}[nosep, leftmargin=*]
    \item Score 1 if relatively minor injury, ambulatory, and stable
    \item Score 2 if potentially life threatening injuries, good capillary refill, stable
    \item Score 3 if airway, breathing, circulation compromises 
    \item Score 4 if unresponsive
\end{itemize}
}

\topic{Essays (ASAP Set 1)}

\tealbox{
\textbf{Prompt: Seed}

Score essay persuasiveness on a scale of 1–6. Students were asked to write a letter to their local newspaper about the effects of computers on people.
}

\tealbox{
\textbf{GEPA-Optimized Prompt: Examples}

You are tasked with grading the persuasiveness of student essays written as letters to a local newspaper, discussing the effects of computers on people. For each student essay provided, follow these steps:

\begin{enumerate}[leftmargin=*]
    \item Read the essay carefully.
    \item Evaluate ONLY the persuasiveness of the essay, not grammar, spelling, or general writing quality, unless these directly impact persuasiveness.
    \item Use the following fixed scale:
    
    1 = Very limited persuasiveness. Argument is minimal, unclear, or nonexistent; little to no support or development.
    
    2 = Weak persuasiveness. Position is stated but lacks support, relevant reasoning, or development.
    
    3 = Some persuasiveness. Position is present with some supporting reasons or evidence, but uneven or limited development.
    
    4 = Adequate persuasiveness. Clear position, sufficient supporting reasons or examples, reasonable organization; may lack sophistication or deep development.
    
    5 = Strong persuasiveness. Well-developed position, relevant and well-chosen support, clear organization, and effective appeals.
    
    6 = Excellent persuasiveness. Highly developed argument, thorough and convincing support, clear and compelling organization and style.
\end{enumerate}
}

\tealbox{
\textbf{GEPA-Optimized Prompt: Distribution}

You are given a task to score student essays for persuasiveness on a fixed integer scale from 1 to 6. Your response MUST be the integer score only—no explanations, justifications, or additional text.

The scoring should be based on the persuasiveness of the essay as demonstrated in the student's response to a prompt, such as writing a letter to their local newspaper about the effects of computers on people. You should evaluate the essay holistically, focusing on the clarity of the position, the coherence, organization and development of arguments, the use of examples or evidence, and overall effectiveness in persuading the reader.

It is critical to adhere to the expected score distribution across all scored items, which should approximately be:
\begin{itemize}[nosep, leftmargin=*]
    \item Score 1: $\sim2\%$
    \item Score 2: $\sim2\%$
    \item Score 3: $\sim10\%$
    \item Score 4: $\sim52\%$
    \item Score 5: $\sim28\%$
    \item Score 6: $\sim6\%$
\end{itemize}

This means the majority of essays should be scored a 4, and only a small minority should be scored at the extreme low (1 or 2) or high (5 or 6) ends of the scale. Calibrate your scoring decisions so that, if applied uniformly, the scores would roughly result in this distribution.

When analyzing the essay, consider the following:
\begin{itemize}[nosep, leftmargin=*]
    \item Does the writer clearly take a position on the topic?
    \item Does the essay include relevant reasons and examples supporting that position?
    \item Is the argument logical, unified, and organized?
    \item How well does the essay address or acknowledge counterpoints?
    \item Do language errors or awkward phrasing significantly detract from persuasiveness, or is meaning generally clear?
    \item Is development limited, superficial, repetitive, or basic (suggestive of a 3 or 4), or is it detailed, well-structured, and sophisticated (suggestive of a 5 or 6)?
\end{itemize}

}

\tealbox{
\textbf{GEPA-Optimized Prompt: Rules}

You are tasked with grading written responses based on a specified scoring rubric. Your goal is to assign a score to each item (response) using only the details provided in the scoring rubric from the task description.


Do NOT simply assign mid-range or "safe" scores; make clear distinctions between rubric levels. If the response fits criteria only for lower rubric levels (minimal or under-developed support, unclear position, etc.), score at the lower end. Only award high scores for responses that clearly and thoughtfully meet the highest-level criteria specified by the rubric.

\medskip
Scoring Rubric:
\begin{itemize}[nosep, leftmargin=*]
    \item Score 1: Undeveloped, takes a position with minimal support.
    \item Score 2: Under-developed, may or may not take a position.
    \item Score 3: Minimally-developed, takes a position, but with inadequate support/details.
    \item Score 4: Somewhat-developed, takes a position, provides adequate support.
    \item Score 5: Developed, clear position, reasonably persuasive support.
    \item Score 6: Well-developed, clear and thoughtful position, persuasive.
\end{itemize}
}

\tealbox{
\textbf{\tool: Rules}
\begin{itemize}[nosep, leftmargin=*]
    \item Score 1 if not persuasive, no clear reasoning
    \item Score 2 if some persuasive arguments, but no clear reasoning
    \item Score 3 if addresses positive side or addresses negative side but provides minimal relevant arguments
    \item Score 4 if persuasive with adequate relevant arguments
    \item Score 5 if persuasive, clear position, addresses both sides, and sufficient revenant examples
    \item Score 6 if persuasive, clear position, addresses both sides, and has strong arguments
\end{itemize}
}

\topic{\textsc{Wands} (Product Relevance)}

\tealbox{
\textbf{Prompt: Seed}

Score the relevance of the given product to the search query on a scale of 0 to 2, where 0 is completely irrelevant and 2 is an exact match.
}

\tealbox{
\textbf{GEPA-Optimized Prompt: Examples}

Your task is to evaluate the relevance of a product to the given search query: ``upholstered bed'', in the home furnishing domain, given the product name, and product description. Assign an integer relevance score between 0 to 2 based on how well the product matches the intent and specificity of the query:

\begin{itemize}[nosep, leftmargin=*]
  \item If the product is the exact kind specified in the query, assign the maximum score.
  \item If the product is not the type of furniture the query asks for, or is unrelated, assign the minimum score, even if there are some similarities (e.g., both are upholstered, but one is a bed and the other is a chair).
  \item If the product somewhat matches the query but is not an exact match (e.g., it is a bed, but not upholstered as required), assign an intermediate score.
  \item Use the product name and description to determine the category of furniture and the specific features required by the query (e.g., ``upholstered'' means the product must have fabric or padding covering parts like the frame or headboard); exclude products that do not fit the primary function or category in the query, even if named similarly.
\end{itemize}
}

\tealbox{
\textbf{\tool: Rules}
\begin{itemize}[nosep, leftmargin=*]
    \item Score 0 if not upholstered and not a bed
    \item Score 1 if a bed but not upholstered, or if upholstered and a different furniture or non-traditional bed
    \item Score 2 if upholstered and a bed
\end{itemize}}

\tealbox{
\textbf{GEPA-Optimized Prompt: Distribution}

You are given a task where you must score the relevance of a product listing to a given search query on a fixed integer scale, based on the query, the product name, and the product description. Output the integer relevance score only, with no reasoning or explanation, following all instructions and calibration guidelines provided below.

\begin{itemize}[nosep, leftmargin=*]
  \item You are judging whether a product is what the user intends. If the query is ``upholstered bed'', only actual upholstered bed products (e.g., bed frames with upholstered headboards/footboards) should be scored as exact matches---not beds in general, sofa beds, or bedding; other related products should be considered less relevant or irrelevant.
  \item Strictly adhere to the provided target score distribution: Score 0 (completely irrelevant): ${\sim}46\%$; Score 1 (partially relevant): ${\sim}30\%$; Score 2 (exact match): ${\sim}24\%$. Use this calibration to avoid overassigning high scores: reserve the highest score only for listings that are an exact match for the query, and use lower scores more frequently.
  \item Take into account both the product name and the description, but the absence of a description is not in itself a reason to award a higher score---in such cases, rely heavily on the product name.
\end{itemize}
}

\tealbox{
\textbf{GEPA-Optimized Prompt: Rules}

You are provided with a scoring rubric, a search query, and details about a product (its name and description). Determine how well the described product matches the intention of the query, and assign an integer score strictly according to the rubric.

\textbf{Domain Rules:}
\begin{itemize}[nosep, leftmargin=*]
  \item ``Upholstered bed'' specifically means a bed (not a sofa bed, daybed, or other convertible) with a frame or headboard that is padded and covered with material. Additional features (e.g., storage, platform style, tufted headboards) do not detract from a match unless they alter the product's primary identity.
  \item A sofa bed or armless sofa bed is not an upholstered bed; it is a partially related product and should not be scored as a direct match.
  \item Beds that explicitly describe themselves as ``upholstered'' in either title or description are an exact match for the query ``upholstered bed''.
  \item If the product fails to mention being a bed, or is clearly a different main type of furniture, it is irrelevant.
\end{itemize}

Be strict: for partial matches, the product must be clearly somewhat related (e.g., an upholstered daybed or a sofa bed for the ``upholstered bed'' query); for exact matches, the product must explicitly match the query and not merely be similar or convertible. Provide only the integer score, with no reasoning or explanation.}

\section{Reproducibility Across Models}
\label{appendix:model-variance-exps}

\add{\tool uses LLM outputs for two stages of the comparative scoring algorithm (Algorithm \ref{alg:attune}): \ttt{Pairwise\_Compare} and \ttt{Discover\_Criteria}. Here, we evaluate whether these judgments are reproducible across models and whether \tool's outputs reflect properties of the task and data rather than model-specific idiosyncrasies. 
}

\topic{Setup} \add{We drew a stratified sample of 100 records each from ASAP and WANDS datasets in our technical evaluation (\Cref{sec:tech-eval}) and all 86 records from \textsc{Triage}, and repeatedly performed pairwise comparisons and criteria discovery with four models spanning distinct families and scales: Qwen 3.5 (2B), Gemma 2 (9B), Claude Opus 4.7, and GPT-4.1. For each dataset, all models judged the identical sampled pairs to produce pairwise preference judgments (A wins, B wins, or tie),  criteria sets, and to assign scores.}

\topic{Measures} \add{For pairwise comparisons, we report Cohen's $\kappa$~\cite{cohen1960coefficient} averaged over the six model pairs, and Krippendorff's $\alpha$~\cite{krippendorff2018content} across the four models. For assigned scores, report Cohen's $\kappa$ and quadratic-weighted $\kappa$ (QWK) averaged over the six model pairs to measure agreement for ordinal scores~\cite{williamson2012framework, cohen1968weighted}, and Krippendorff's $\alpha$ across the four models. For criteria discovery, we report Jaccard similarity coefficients~\cite{jaccard1912distribution} to measure the overlap of matched criteria. }

\begin{table}[htp]
  \centering
  \small
  \setlength{\tabcolsep}{4.5pt}
  \caption{\add{Inter-model agreement on pairwise comparisons and score
  assignments, per dataset.}}
  \label{tab:agreement-pairwise-scores}
  \begin{tabular}{lcc ccc}
    \toprule
    & \multicolumn{2}{c}{Pairwise Comparisons} & \multicolumn{3}{c}{Score Assignments} \\
    \cmidrule(lr){2-3} \cmidrule(lr){4-6}
    Dataset & $\kappa$ & $\alpha$ & $\kappa$ & QWK & $\alpha$ \\
    \midrule
    \textsc{Triage} & 0.84 & 0.63 & 0.41 & 0.69 & 0.69 \\
    ASAP            & 0.37 & 0.23 & 0.18 & 0.42 & 0.42 \\
    WANDS           & 0.56 & 0.40 & 0.48 & 0.61 & 0.61 \\
    \midrule
    Mean            & 0.59 & 0.42 & 0.36 & 0.57 & 0.57 \\
    \bottomrule
  \end{tabular}
\end{table}

\topic{Results} \add{\Cref{tab:agreement-pairwise-scores} reports inter-model agreement for pairwise comparisons and score assignments. We learn that inter-model agreement in pairwise comparisons ranges from fair to almost perfect agreement for the three datasets, where the Cohen's $\kappa$ averages 0.59 indicating moderate agreement~\cite{landis1977measurement}. Models show an agreement percentage of 66.2\% across datasets. 
Using \tool's score assignment algorithm also reflects fair to moderate agreement across models, with the quadratic-weighted $\kappa$ (QWK) averaging 0.57. The gap between the exact-match $\kappa$ (0.36) and the QWK indicates that \tool placed records in the same ordering, but the underlying pairwise comparisons led score assignments to differ by one or more score bins. On inspecting the distribution of ties in pairwise comparisons, we learn that model families differ in how many ties they declare, ranging from 5\% (Gemma 2) to 23\% (Claude Opus 4.7). We also observe that inter-model agreements are lowest for the ASAP dataset, reflecting the highly subjective nature of essay grading and a wider scoring scale.}

\add{For discovered criteria, we observe Jaccard similarity coefficients of 0.84, 0.53, and 0.76 for \textsc{Triage}, ASAP, and WANDS respectively. The models primarily exhibited differences in the cardinality of the criteria set based on how finely they decompose criteria. For example, in \textsc{Triage}, Qwen 3.5 reports ``Tachycardic'' and ``Slow pulse'' as separate criteria, while Claude Opus 4.7 resolved these under ``Abnormal Pulse''; and in ASAP, GPT-4.1 reports ``Strong Introduction present'' and ``Strong Conclusion present'' as separate criteria, while Gemma 2 checks for ``Coherent Structure'' only. We also observed that 91\% of all models' criteria were covered by at least one other model across datasets, indicating stable criteria coverage across models.} 
\section{User Study: Supplementary Details}
\label{appendix:summative-details}

\subsubsection*{Participant Backgrounds} We recruited participants with domain expertise in healthcare, education, law, and AI-based evaluation. \Cref{table:participants} summarizes their backgrounds and study datasets.

\begin{table}[htp]
\centering
\caption{Summative study participant backgrounds and datasets.}
\label{table:participants}
\footnotesize
\begin{tabular}{llll}
\toprule
\textbf{ID} & \textbf{Domain, Specialization} & \textbf{Exp (yrs)} & \textbf{Dataset} \\
\midrule
P1 & Healthcare, Emergency Surgeon & 20--30 & \textsc{Triage} \\
P2 & Law, Real Estate Attorney & 10--20 & Lease Red Flags \\
P3 & Education, TA for CS Education & 2--5 & Essays (ASAP Set 1) \\
P4 & Education, TA for CS Education & 1--2 & Essays (ASAP Set 1) \\
P5 & Education, TA for Computing Ethics & 2--5 & Essays (ASAP Set 1) \\
P6 & AI Evaluation, RAG Engineer & 10--20 & Recommendations \\
P7 & Healthcare, Compliance & 20--30 & \textsc{Triage} \\
P8 & AI Evaluation, RAG Researcher & 2--5 & Recommendations \\
\bottomrule
\end{tabular}
\end{table}

\subsubsection*{Study Tasks}
We prepared scoring tasks spanning participant expertise. The first two tasks were carried over from the formative study and technical evaluation (Table~\ref{table:probe-tasks}).

\begin{table}[htp]
\centering
\caption{Datasets and tasks used in the user study.}
\label{table:summative-tasks}
\footnotesize
\begin{tabular}{lp{5cm}}
\toprule
\textbf{Dataset} & \textbf{Task} \\
\midrule
\textsc{Triage}~\cite{kirch2024triage} & Score severity of patient injuries on a scale of 1--4. \\
Essays (ASAP Set 1)~\cite{mathias2018asap} & Score essay persuasiveness on a scale of 1--10. \\
Lease Red Flags~\cite{leivaditi2020lease} & Score rental lease agreements on a scale of 1--5 based on red flags and potential risks. \\
Recommendations & Score retrieved movie recommendations based on personal preference on a scale of 1--10.\\
\bottomrule
\end{tabular}
\end{table}

\subsubsection*{Post-study questionnaire}
Participants responded to the following items on a 5-point Likert scale after their study session:

\begin{enumerate}[label=\textbf{Q\arabic*.}, leftmargin=*]
  \item How would you rate the quality of generated scores?
  \item How helpful was it to annotate samples?
  \item How helpful were the scoring criteria?
  \item How helpful were the scoring rules?
  \item How helpful was it to be able to edit the score distribution?
  \item How helpful was it to be able to provide example scores?
  \item How easy was it to steer the system?
  \item To what extent did you feel in control of the scoring decisions?
\end{enumerate}
\smallskip

\begin{figure}[ht]
    \centering
    \includegraphics[width=1\linewidth]{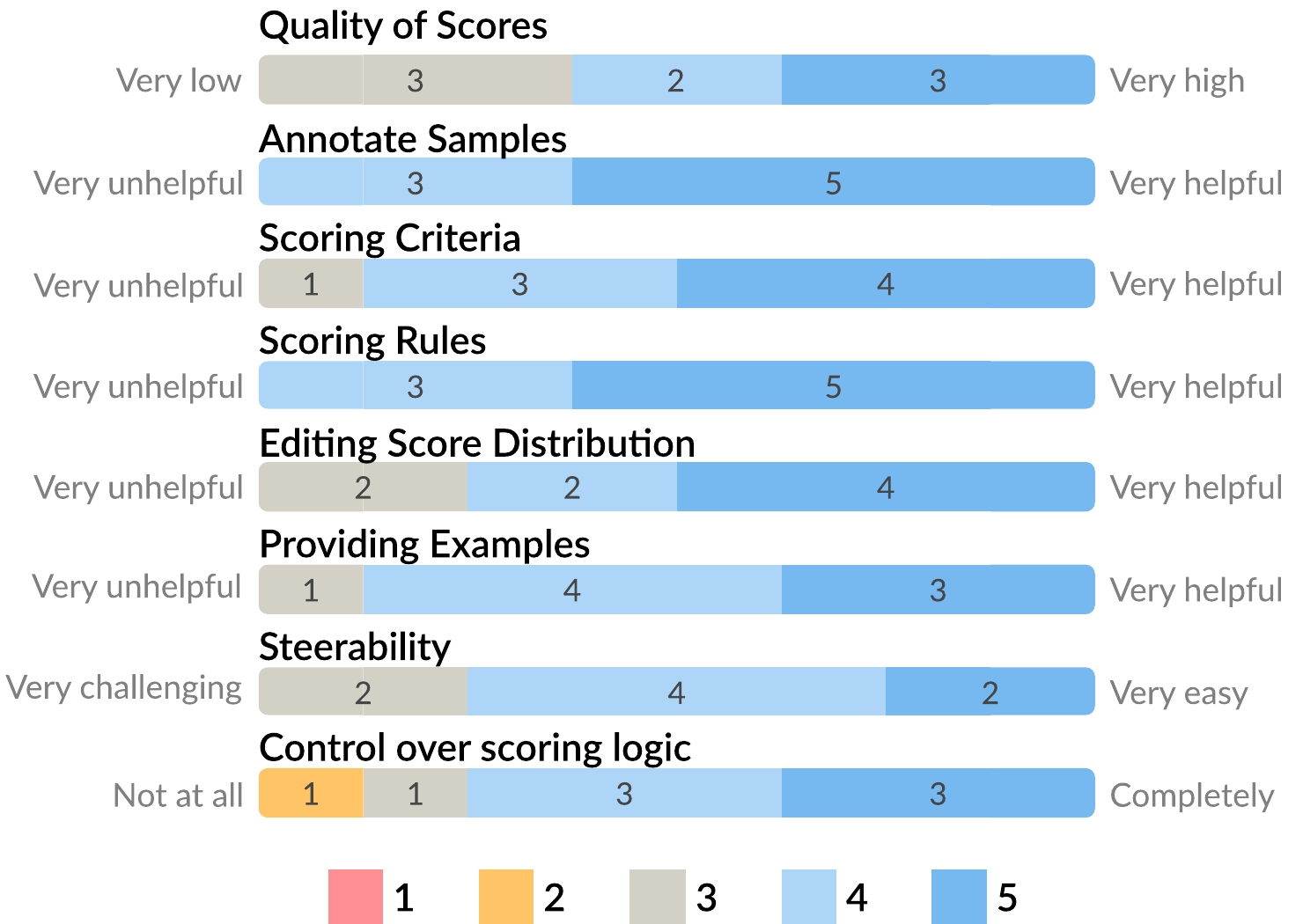}
    \caption{\add{Participant responses to post-study questions}}
    \Description{}
    \label{fig:responses}
\end{figure}

\noindent \add{Figure \ref{fig:responses} summarizes participant responses for each question.}

\end{document}